\documentclass[12pt,a4paper]{article}

\usepackage{geometry}
\usepackage{bbm,yfonts}
\usepackage{epstopdf}
\usepackage{revsymb}
\usepackage[nottoc]{tocbibind} 
\usepackage{xspace}	
\usepackage{mathtools}
\usepackage{graphicx}
\usepackage{ytableau}
\usepackage{pgf,pgfarrows,pgfnodes}
\usepackage{graphics}
\usepackage[colorlinks=true,
			urlcolor=blue,
			linkcolor=black]{hyperref}
\usepackage{tikz}
\usepackage{xparse}
\usepackage{subfigure}
\usepackage{scrextend}
\usepackage{amssymb}
\usepackage{cite}

\usepackage{ifthen}	

\usepackage{titlesec}					
\usepackage{graphicx}
\usepackage{amsmath}
\usepackage{amssymb}
\usepackage{epsfig}
\usepackage{amsthm}
\usepackage{color}
\usepackage{animate}
\def\be{\begin{equation}}
\def\ee{\end{equation}}
\def\arr{\begin{array}{rll}}
\def\ea{\end{array}}
\def\bea{\begin{eqnarray}}
\def\eea{\end{eqnarray}}

\newcommand\scalemath[2]{\scalebox{#1}{\mbox{\ensuremath{\displaystyle #2}}}}

\titleformat{\section}[display]{\normalfont\Large\bfseries}
{CHAPTER \thesection}{0em}{\MakeUppercase} 
\titlespacing*{\section}{0em}{0em}{5em}   

\titleformat{\subsection}{\normalfont\normalsize\bfseries}
{\thesubsection}{1em}{\MakeUppercase} 
\titlespacing*{\subsection}{0em}{0em}{1em}   

\titleformat{\subsubsection}{\normalfont\normalsize\bfseries}
{\thesubsubsection}{1em}{\MakeUppercase} 
\titlespacing*{\subsubsection}{0em}{3em}{2em}   

\newcommand{\chap}[1]{\textbf{Chapter #1}}
\newcommand{\sect}[1]{\textbf{Section #1}}

\newcommand{\beq}{\begin{equation}}
\newcommand{\eeq}{\end{equation}}
\newcommand{\beqarr}{\begin{eqnarray}}
\newcommand{\eeqarr}{\end{eqnarray}}
\newcommand{\ber}{\begin{array}}
	\newcommand{\eer}{\end{array}}

\graphicspath{
	{images/}
	{.}
}

\DeclareMathAlphabet{\pazocal}{OMS}{zplm}{m}{n}

\def\be{\begin{equation}}
\def\ee{\end{equation}}
\def\bea{\begin{eqnarray}}
\def\eea{\end{eqnarray}}

\def\N2{$N{=}2$}

\def\l3{\ell_3}

\newcommand{\neweq}[2]{\ifthenelse{\equal{#1}{*}}{\begin{equation*}#2\end{equation*}}
	{\ifthenelse{\equal{#1}{}}{\begin{equation}#2\end{equation}}
		{\begin{equation}\label{#1}#2\end{equation}}}}	

\begin{document}

\thispagestyle{empty}
\def\thefootnote{\fnsymbol{footnote}}
\vspace*{0.5cm}
\begin{center}
{\large Institute of Radiophysics and Electronics}\\
\vspace*{5cm}
{\bf\Large Mher Davtyan}\\
\vspace{1cm}
{\bf\Large The role of symmetrical inhomogeneity in problems of electromagnetic wave propagation}\\
\vspace{1cm}
{\Large PhD Thesis}\\
\vspace{1cm}
{\large 01.04.03 - Radiophysics}\\
{\large for PhD in physics and mathematics}\\
\vspace{2cm}
{\qquad\qquad\qquad\qquad\qquad\qquad\large Advisor: Zhirayr Gevorkian, Doctor of Science}
\vspace{3cm}
{\large Yerevan-2022}
\end{center}

\newpage

\tableofcontents

\newpage

\section{Introduction}

Media with symmetric inhomogeneity have been of great interest due to numerous effects which occur when electromagnetic waves propagate through such media. In general, inhomogeneity of a certain medium means that the refraction index of the medium is not constant and has some sort of spatial dependency. These types of media are called gradient index (GRIN) media. Different optical effects related to GRIN media are described by GRIN optics. GRIN optics examines different phenomena related to the gradient of the refractive index of the medium and/or the material. It is known that one important advantage of GRIN materials is that those can be used to build optical systems for imaging without aberrations which are common for traditional optical systems like spherical homogenous lenses. GRIN materials are classified into three major categories – axial, cylindrical and spherical. In axial GRIN profiles the refractive index changes along the optical axis of the system. Cylindrical refractive profile is a system consisting of concentric cylinders with surfaces with constant refraction index. Spherical profiles, by analogy, are symmetric relative to point source. In spherical profiles surfaces of concentric spheres have constant refraction index. Electromagnetic waves propagating in such materials result in many effects which are being exploited to produce lenses, optical fibers and other devices.
By symmetric inhomogeneity we refer to the case when the inhomogeneity of the medium or material is not an arbitrary function of position but has some sort of symmetry. In this dissertation our main focus is on two types of inhomogenous media. Firstly, we examine photonic crystals which are optical structures where the refractive index changes periodically. The other type of inhomogeneity that is examined is the case when the refractive index of a medium is a continuous function of position. Moreover, central focus is given to profiles possessing different symmetries. Specifically, we investigate the effects occurring in Maxwell Fish eye profile which has spherical symmetry as well as an extended symmetry which will be thoroughly studied in Chapters 3 and 4.

Photonic crystal is a system with periodical dielectric permittivity and has many interesting
properties regarding transmission and polarization change of incident electromagnetic wave. Recent developments in material science have made possible the fabrication of photonic crystals that allow the observation of many peculiar effects \cite{Joan08}, including perfect reflector for all polarizations over a wide selectable spectrum \cite{john,oscar}, optical Hall effect \cite{OMN2004}, unidirectional scattering \cite{haldane08} with broken time reversal symmetry, the propagation of optical beams without spatial spreading and polarization rotation effects in dilute photonic crystal which were exmained in \cite{ghgc17}.
In general, it is accepted that in geometrical optics approximation polarization change is negligible. Conversely, in the opposite limit, when the scale of characteristic inhomogeneity is of same order as that of the wavelength - rotation of polarization becomes much more noticeable. The problem of polarization change is particularly important because of its applications in polarization controlling devices and when there is a need to manipulate the polarization of incidence light \cite{solli2004,li2001, solli2003}. Particularly, in the above-mantioned paper, it is experimentally observed that there appears a drastic change of polarization in photonic crystal which has inhomogeneity perpendicular to the propagation direction of the wave. In \sect{\ref{sec:capsize-experiment}} we briefly present the expermental set-up that was used to detect the change of polarization. In the following two sections it is presented the developed theory which is based on Maxwell's equations with two dimensional inhomogeneous permittivity. In \cite{gevdav19} we have presented the theory for the TE waves and obtained a good correspondence between the theoretical and experimental frequencies for which the resonance polarization change was observed.
Under certain conditions dilute photonic crystals are also argued to cause a light straightening effect \cite{ggc19}. It is found that such effect is caused by the formation of localized states in transversal motion. It is argued that these states decrease mobility in the transverse direction 
and force the light to be straightened.

Starting from the second chapter we mainly focus on the other type of the inhomogeneity that was noted earlier. That is we will examine various phenomena that arise when electromagnetic wave propagates through different kinds of continuous  refractive media. Particularly, we have considered refractive profiles (Maxwell fish eye, Luneburg lense) posessing spherical symmetries which are of high scientific interest due to their applications in well-known phenomena of perfect imaging and cloaking. Cloaking phenomena have attracted a reasonable amount of interest since Pendry \cite{Pendry06} and Leonhardt \cite{Leonhardt06} assumed that an object coated by certain inhomogeneous shell becomes invisible to electromagnetic waves. Different mechanisms of cloaking have been suggested since that time, among which are anisotropic metamaterial shells \cite{Pendry06}, conformal mapping in two-dimensional systems \cite{Leonhardt06}, complementary media \cite{lai09}, etc.
Transformation optics \cite{Pendry06} is the most common approach, where the dielectric permittivity and magnetic permeability tensors are specific coordinate-dependent functions. However, this approach is difficult to implement in the optical field, since it is problematic to find metamaterials with the necessary magnetic properties \cite{zhou05}. In the cases when the photon wavelength is much smaller than the characteristic size of inhomogeneity, geometrical optics approximation is justified \cite{Sun08}-\cite{choi14}. In the conformal mapping method of cloaking when geometrical optics approximation is used,  closed ray trajectories are of significant importance \cite{Leonhardt06}. The scheme that Leonhardt proposed uses optical conformal maps where  a dielectric medium conformally maps physical space onto Riemann sheets. The idea is to send all rays that have passed through the branch cut onto the interior sheet back to the cut at precisely the same location and in the same direction in which they entered \cite{Leonhardt06}. The latter illustrates that potentials (refractive profiles) which result to closed ray trajectories are particularly important in achieving the cloaking phenomenon. Moreover, these trajectories determine the size and shape of the cloaked area. Other that that, Maxwell's fish eye has also applications in quantum optics. While investigating quantum optical properties of MFE, it was shown that such a system mediates effectively infinite-range dipole-dipole interactions between atomic qubits, which can be used to entangle multiple pairs of distant qubits \cite{perczel}. MFE was also used as a possible mean for spin waves focusing \cite{haitao}. Using numerical simulations it was shown that sub-wavelength focusing can be achieved by means of MFE with appropriate chosen parameters. Furthermore, simulation results also prove that the focusing properties of MFE can be tuned by external magnetic field, so the MFE would be promising device for spin wave focusing and wavefront manipulation \cite{haitao}.
Maxwell fish eye potential energy profile was also considered in graphene quantum dot. It was shown that all the electron trajectories are closed circles that are classified by angular momentum and an additional integral of motion. Moreover, given that the lower Dirac zone is completely filled, a universal value for Hall conductivity is found \cite{gevorkian-graphene}.

Note that in this section of the dissertation light polarization is mostly neglected.

In Chapter 4 we examine the propagation of polarized light in the medium with Maxwell fish eye refraction index profile. The chapter is organized as follows. In \autoref{sec:fish-eye-polarized-scalar-waves}, we describe the Hamiltonian formulation of the geometric optical system given by the action \eqref{gactions2}. We also present some other textbook facts on the duality between Coulomb and free-particle systems on a (pseudo)sphere which allow us to relate the Maxwell fish eye and Coulomb profiles and will be used in our further consideration. In \autoref{sec:fish-eye-polarized-scalar-waves}, we present the Hamiltonian formalism for the polarized light propagating in an optical medium and propose the general scheme of the deformation of an isotropic refraction index profile which allows us to restore the initial symmetries after the inclusion of polarization. In \autoref{sec:fish-eye-polarized-spin-inclusion}, we use the proposed scheme for the construction of a “polarized Maxwell fish eye” profile which inherits all the symmetries of the original profile when light polarization is taken into account. We present the explicit expressions for the symmetry generators of the corresponding Hamiltonian system and find the expressions of the Casimirs of their symmetry algebra. In \autoref{sec:fish-eye-polarized-trajectories} the explicit expressions for the trajectories of polarized light are presented. It is shown that these trajectories are no longer orthogonal to the angular momentum but turn to a fixed angle relative to it. Despite deviations from circles, these trajectories remain closed. We show that light polarization violates the additional symmetries of the medium so that ray trajectories no longer remain closed. Afterwards we suggest a modified, polarization-dependent  Maxwell fish eye refraction profile which restores all the symmetries of the initial profile and yields closed trajectories of polarized light. Explicit expressions for the polarization-dependent integrals of motion and the solutions of corresponding ray trajectories are also presented.

In Chapter 5 we consider problems related to light generation in mediums with different types of refraction index profiles. So far most of the attention was paid to light propagation in such systems. Other very interesting topic which also has a big potential in terms of real life applications is light generation in such profiles. Within the scope of this dissertation, light generation mechanisms in a medium with highly symmetric Maxwell fish eye refractive index profile are investigated. Under the term light generation, we refer to radiation properties of the Maxwell fish eye when a charged particle interacts with the refraction profile by penetrating through it. In general there are three main types of emitted radiation when a charged particle penetrates some medium. Those are Cherenkov radiation, transition radiation and finally the so called diffraction radiation. Radiation intensity and spectrum are primary objectives of investigation in any type of radiation problem. Apart from those, angular distribution of the radiation intensity is another important property which is examined in this section. Whether the radiation is isotropic or it is highly directed towards some direction. 
In this chapter radiation from a charged particle moving in a medium with Maxwell fish eye refraction index profile is considered. Maxwell’s equations are used to describe the electromagnetic fields in an inhomogeneous medium taking into account the moving charge as an external source. Other than that, well-known method of Green’s function is used to find out the expression for radiation potential $A$ and corresponding electric and magnetic fields. Generally, Green's function is an integral kernel that can be used to solve many types of differential equations including inhomogeneous partial differential equations \cite{green-wolfram}. Also, we have constructed the well-known Poynting’s vector which will be used to derive the radiation intensity. Whenever appropriate, numerical estimations of integrals are completed. We have shown that the radiation spectrum has a discrete character. The main emitted wavelength is proportional to the refractive profile's radius and has a dipole character in a regular medium. Cherenkov like threshold velocity is established. A cardinal rearrangement of angular distribution in a lossless medium is predicted. This behavior is caused by the total internal reflection in a lossless medium as apposed to photons' attenuated total  reflection in the regular medium. Lossless medium ensures that both directed and monochromatic emission can serve as a light source in the corresponding regions.

\section[Polarization rotation in photonic crystal]
    {Polarization rotation in photonic crystal}
\label{sec:pol_rot}

\subsection{Introduction}
\label{sec:capsize-intro}
Rotation of polarization of TE waves in one dimensional dilute photonic crystal is theoretically
investigated. Resonance character of polarization rotation is revealed. It is shown that rotation angle at
resonant frequencies acquires only discrete values. Results of theory are compared with the experiment.

In \sect{\ref{sec:capsize-experiment}} we describe the recent experimental results obtained in dilute photonic crystal where resonance polarization change is observed. 

In \sect{\ref{sec:capsize-theory-tm}} we present the theory that was introduced in \cite{ghgc17}. In \sect{\ref{sec:capsize-theory-te}} it is presented the reformulation of the same theory using TE waves. In the following sections it is examined the spectrum of the photonic crystal as well as comparison with the actual experimental results is completed. As we noted earlier good correspondence between the TE theory and the experiment is observed.

\subsection{Experiment}
\label{sec:capsize-experiment}
\begin{figure}
 \begin{center}
\includegraphics[width=16.0cm]{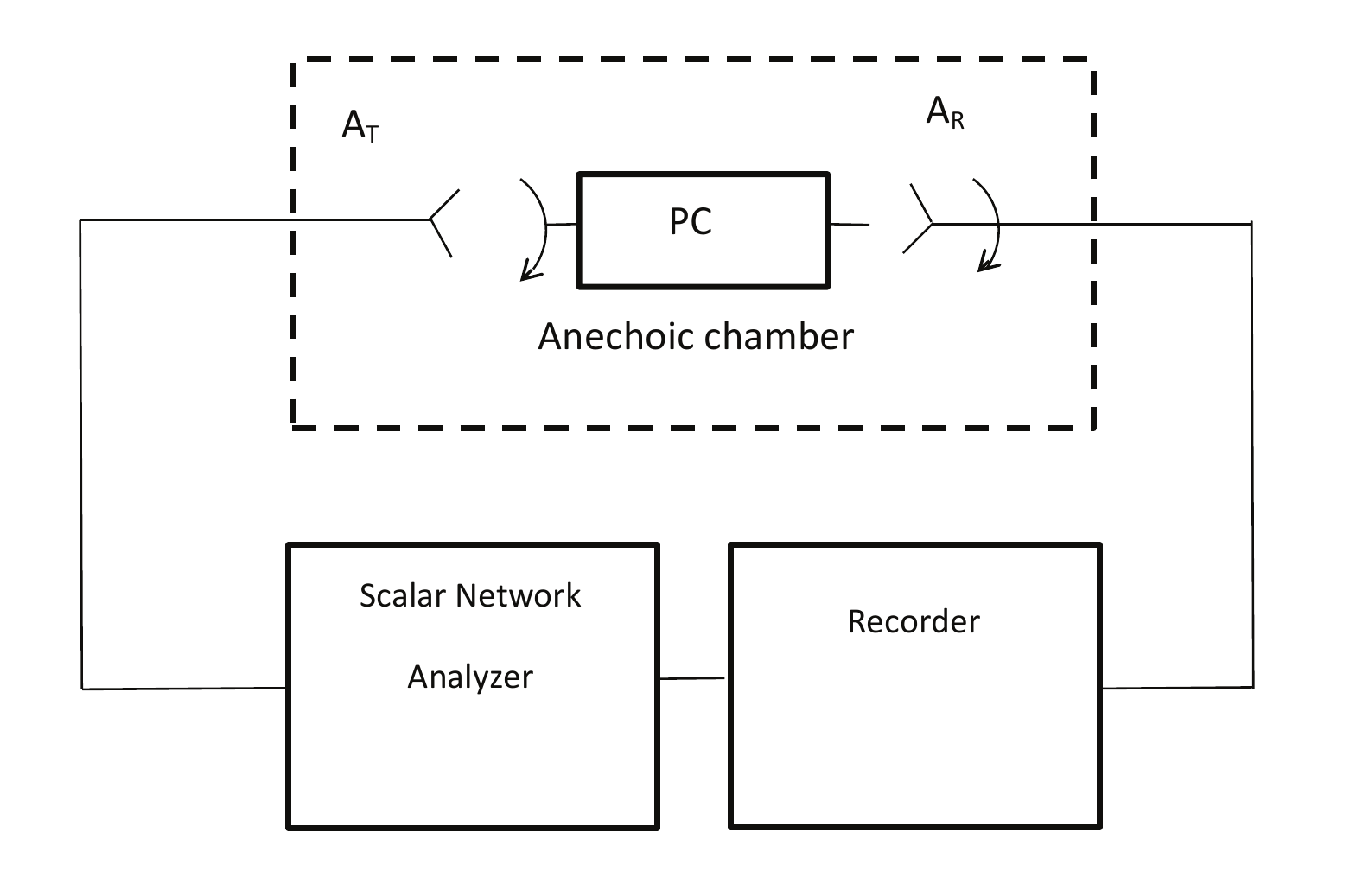}
\caption{Schematic of the experimental setup. Waves propagate through the dilute PC structure.
The arrangement is used to detect the frequency dependence of the normalized intensity $I$.}
\label{pol-rot-setup}
\end{center}
\end{figure}

\begin{figure}
\begin{center}
\includegraphics[width=16.0cm]{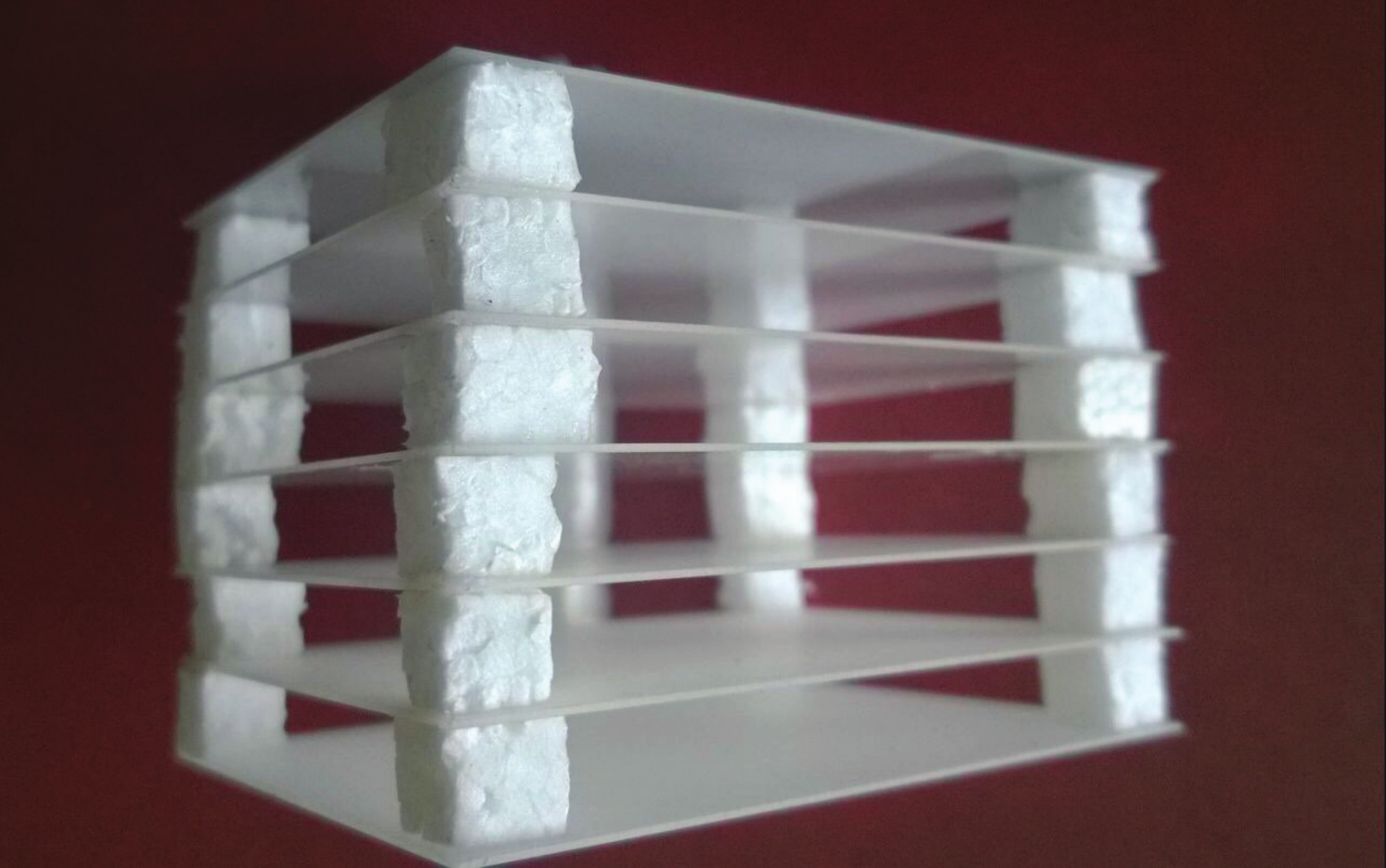}
\caption{One dimensional photonic crystal. Photograph of the 1d PC used in the experiment.
It is formed from $7$ alumina ceramic plates with $\varepsilon=10$, thickness $b=0.5$ mm, sizes
$60\times 96$ mm$^2$ and $a=10$ mm distance from each other.}
\label{pol-rot-1dpc}
\end{center}
\end{figure}

To study capsize effect in 1d PC system, we have performed an experiment where electromagnetic waves propagate through the dilute structure. The schematic setup of the experiment is shown in Fig. \ref{pol-rot-setup}.
It consists of transmitter and receiver antennas, an anechoic chamber with photonic crystal, scalar network analyzer and recorder. One-dimensional photonic crystal is formed from $7$ alumina ceramic plates with permittivity $\varepsilon=10$, thickness $b=0.5$ mm and sizes $60\times 96$ mm$^2$.
They are parallel substituted along the $x-$axis at the distance $a=10$ mm from each other. The structure is supported by the low loss and low dielectric permittivity $\varepsilon=1.07$ foam layers, see Fig. \ref{pol-rot-1dpc}.
The measurements were carried out in the frequency range $10\div 140$ GHz (wavenumber $k_0=0.21\div 2.93$ mm$^{-1})$ using set of scalar network analyzers. Open end rectangular waveguides of corresponding frequency range are used as a transmitting and receiving antennas. Such a choice of antennas allow high level cross-polarization isolation smaller than $10^{-3}$
in their mutual orthogonal location. The PC and the receiving antenna can be rotated in the $xy-$plane, where the electric field vector ${\bf E}(\bf H)$ is located.

We remind that the wave incidence direction is $z-$axis and plates are located in the $yz$-plane. The PC and the antennas were placed in an anechoic chamber to avoid the influence of unwanted external signals. The distances between them were chosen large enough that far field approximation is applicable. The frequency dependence of transmission coefficient was displayed on the recorder.

The results of experiment with one dimensional photonic crystal are presented in Fig. 3. In the upper panel (black line) the frequency dependence of transmission coefficient $I$ (normalized intensity) is shown provided that wave's incidence direction ($x-$axis) is perpendicular to the plates. Due to the periodicity in $x-$direction we expect to have frequency bands over which the propagation of the waves is forbidden, that is the transmission coefficient becomes zero. The panel indicates that at wave numbers above 0.48 mm$^{-1}$ the forbidden band begins to form.
In the middle and lower panels (red and blue lines, respectively) the wave incidence direction is $z-$axis, as discussed below. However, the plots differ from each other by the direction of receiving antenna: in the middle panel it is orthogonal to transmitter antenna ($y-$axis) and in the lower panel it is parallel to transmitter, that is directed in $x-$axis. The maxima in the middle panel corresponds to the resonance wavelengths $v_{1d}L=\pi/2+\pi n$(see theory below) for which polarization change the direction from $\pi/4$ (initial linear polarization angle) to $\pi/2$. In the lower panel (blue line) for the same wavelengths one has minimums instead of maximums. This shift of the intensity distribution is due to the change of receiver's direction: the latter now is pointed out on  $x-$direction, that is parallel to the transmitter polarization.To verify the preceding discussion on the PR in 1d PC around resonance wavelengths $v_{1d}L=\pi/2+\pi n$, one needs to evaluate $v_{1d}$. Using equation  (\ref{spl1d}) we have calculated $v_{1d}(k_0)$ for the resonant wavelength $k_0=0.49$mm$^{-1}$ and for $\cos v_{1d}(k_0=0.49)L$ we get $-0.008$ ($L=96mm$) which is very close to $0$, as was predicted. Note the logarithmic scale used in the experimental plot Fig. 3. This means that the capsize phenomena takes place with very high accuracy $10^{-3}$.

\begin{figure}
\begin{center}
\includegraphics[width=16cm]{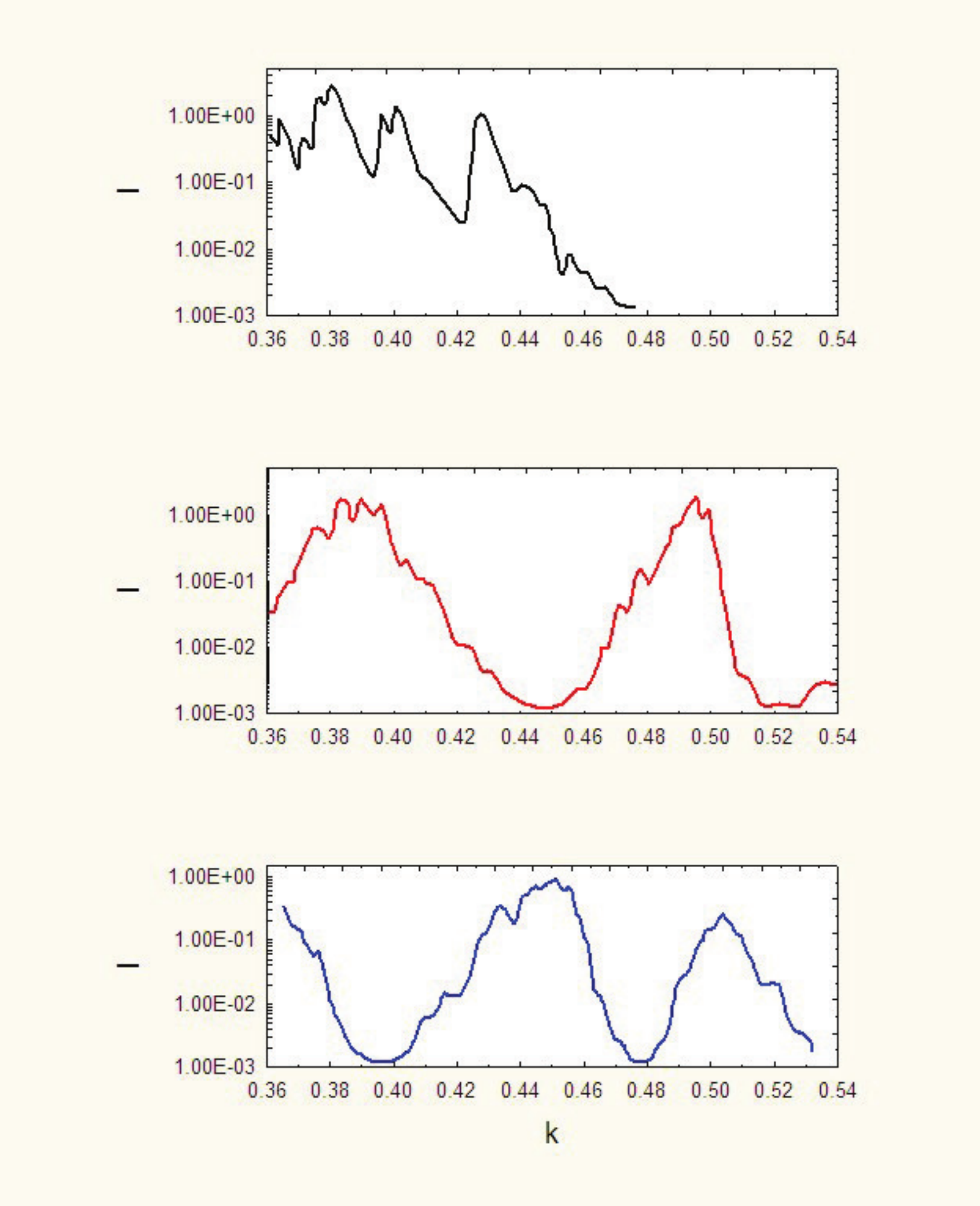}
\caption{Measured intensity for different wave's incidences.
The upper panel (black line) shows intensity when the wave normally falls on the parallel
plates that make one-dimensional photonic crystal. It indicates that at wave numbers above 0.48 mm$^{-1}$ the forbidden band begins to form in the frequencies range of interest. In the middle and lower panels (red and blue lines, respectively) the wavevector is parallel
to the plates.  However, the receiver is pointed out in $y-$direction (red line) and in
$x-$direction (blue line). The incident wavevector is on $z-$axis and the photonic crystal
is rotated on $\pi/4$ in the $xy-$plane -therefore initial polarization is at $\pi/4$.}
\label{fig.3}
\end{center}
\end{figure}

As we mentioned above, at $\theta_0=\pi/4$ and for resonant frequencies $v_{1d}L=\pi/2+\pi n$, the transmitted wave is linearly polarized but directed on $y-$axis. This means that the theoretical imaginary part of rotation angle that describes the ellipsity (ratio of ellipse axes) should become infinite.  However, in general, for arbitrary $\theta_0$ there is a finite ellipticity and the imaginary portion ${\rm Im} \theta(L)$ is also finite. In Fig. 4 we compare experimentally measured ellipticity at resonance wavenumber $k_0=0.49$ mm$^{-1}$ (red line) with the theoretical ellipticity (blue line) (see below, equation  (\ref{im1d})).

\subsection{Theory: TM wave}
\label{sec:capsize-theory-tm}
In achieving a better understanding of the experimentally observed capsize effect, discussed in the previous section, we develop an analytical approach that predicts the capsize of polarization in 1d and 2d DPCs.

Let us consider a dielectric medium with periodic array of two dimensional holes.
The inhomogeneity of the medium in $xy-$plain will be taken into account through a spatially dependent two dimensional dielectric permittivity $\varepsilon(x,y)$. Suppose a plane wave enters the medium from the $z<0$ half-space at normal incidence. Using Maxwell equations one can obtain a wave equation for $\bf H$ in the form
\begin{eqnarray}\label{sqeq}
\nabla^2{\bf H}(x,y,z)+k_0^2\varepsilon(x,y){\bf H}(x,y,z)+\\
\nonumber+\frac{\nabla\varepsilon(x,y)}{\varepsilon}\times{\bf \nabla}\times {\bf H}(x,y,z)=0,
\end{eqnarray}
where $k_0=\omega/c$ is the wave number corresponding to the angular frequency $\omega$ of the incident wave.
The last term in l.h.s. of equation (\ref{sqeq}) accounts for the contribution of the dielectric permittivity inhomogeneity and, as we will see below, introduces a correction to the band structure of the two-dimensional Bloch states. This correction creates the dynamical phase shift between the waves propagating in the orthogonal directions in the $xy-$plain and finally leads to capsizing of the initial polarization of $\bf H$.

Before proceeding further, we would like to mention that similarly to equation  (\ref{sqeq}) for $ \bf H$, one can write a wave equation for the electric field $\bf E$. However, the wave equation for $\bf E$ will contain en extra term ${\bf E}\nabla({\bf \nabla}\varepsilon/ \varepsilon)$ that does not affect the initial polarization. For this reason and for simplicity in what follows, we study only the peculiar rotation of the $\bf H$ in $xy-$plain caused for the dielectric permittivity inhomogeneity.

Following Refs. \cite{FF79,Lag,GGC16}, we seek the solution for the propagation in the system wave as a product of a fast and a slowly varying, $\bf\tilde{H}(x,y,z)$, function on a wave incident $z-$direction
\begin{equation}
\scalemath{1}{
{\bf H}(x,y,z)=e^{ik_0z} {\bf \tilde{H}}(x,y,z).
}
\label{slfas}
\end{equation}
Substituting equation (\ref{slfas}) into equation (\ref{sqeq}) and neglecting the second derivative of $\tilde{H}$ with respect to $z$ one gets in the parabolic equation approximation
\begin{equation}
\scalemath{1}{
i\frac{d\tilde{{\bf H}}}{dz}=\hat{H}(x,y)\tilde{{\bf H}},
}
\label{sreq1}
\end{equation}
where $\tilde{{\bf H}}\equiv\binom{\tilde{H}_x}{\tilde{H}_y}$ is a spinor and $\hat{H}=\hat{H}_0+\hat{V}$ is a $2\times 2$ matrix.

The explicit form of
$\hat{H}_0$ is
\begin{equation}
\scalemath{1}{
  \hat{H}_0=
\begin{pmatrix}
 -\frac{1}{2k_0}\nabla_\bot^2+\frac{k_0}{2}(1-\varepsilon(x,y)) & 0 \\
 0 & -\frac{1}{2k_0}\nabla_\bot^2+\frac{k_0}{2}(1-\varepsilon(x,y))  \label{tem}
\end{pmatrix}
}
\end{equation}
where $\nabla_\bot^2\equiv \partial^2/\partial x^2+\partial^2/\partial y^2$
and the inhomogeneity term is given by
\begin{equation}\label{inhom}
\scalemath{1}{
 \hat{V}=\frac{1}{2k_0\varepsilon}
\begin{pmatrix}
-\frac{\partial \varepsilon}{\partial y}\frac{\partial}{\partial y} & \frac{\partial \varepsilon}{\partial y}\frac{\partial}{\partial x}\\
\frac{\partial \varepsilon}{\partial x}\frac{\partial}{\partial y} & -\frac{\partial \varepsilon}{\partial x}\frac{\partial}{\partial x}
\end{pmatrix}.
}
\end{equation}
Let us stress that ignoring the second
derivative of $\tilde{H}$ with respect to $z$ in equation  (\ref{sreq1}) is justified by the fact that in a dilute system with a small fraction of dielectric (or metal) the light propagates in $z-$direction mostly forwardly, without back scattering.

So the problem of propagation of electromagnetic waves in a dilute dielectric medium with periodic array of two dimensional holes is reduced to a quantum motion of a particle with mass $k_0$ in a $2d$ periodic potential $V(x,y)=\frac{k_0}{2}\left(1-\varepsilon(x,y)\right)$ with inhomogeneity term $\hat{V}$, where spatial coordinate $z$ plays the role of time in a Schr{\"o}dinger-like
equation. In this approach, if the initial parameters at $z=0$ are known (for example the incoming wave front), then one can find them at $z>0$.  Such a mapping has been performed previously \cite{Lag,GGC16} and made it possible to study the propagation of light through a semi-infinite medium with transverse disorder \cite{Lag} and through  a dilute metal with two-dimensional hole arrays \cite{GGC16}.

It can furthermore be shown that if one associates $\tilde{H}_{x,y}$ components with spin components then the term $\hat{V}$ reminds Rashba spin-orbit interaction (see for a review Ref. \cite{Rashba15}, for photon spin-orbit interaction  \cite{Libzel92} and the review\cite{Bliokh15}). The remarkable similarity between $\hat{V}$, equation  (\ref{inhom}), and a Rashba spin-orbital interaction term
allow us to stress that both are closely related phenomena. The term $\hat{V}$ will introduce a correction to the band structure of the two-dimensional Bloch states, creates the dynamical phase shift between the waves propagating in the $x-$ and $y-$directions and finally leads to rotation of the initial polarization.
Hence, the calculation of the spectrum of Hamiltonian $\hat{H}(x,y)=\hat{H}_0+\hat{V}$ could provide a sensitive means to analyze polarization effects in dilute photonic crystal and may be used as a starting point to evaluate the wave transmission coefficient at $z-$direction.

To find the band structure let us represent the solution of equation (\ref{sreq1}) through the eigenfunctions of the Hamiltonian equations  (\ref{tem}) and (\ref{inhom}),
\begin{equation}
\scalemath{1}{
\tilde{{\bf H}}(x,y,z)=\sum_nc_ne^{-iE_nz}{\bf h}_n,
}
\label{eig}
\end{equation}
and
\begin{equation}
\scalemath{1}{
\hat{H}{\bf h}_n(x,y)=E_n{\bf h}_n(x,y).
}
\label{eigfun}
\end{equation}

Finally, substitution of equation (\ref{eig}) into equation (\ref{slfas}) yields the solution of Maxwell equation
\begin{equation}
\scalemath{1}{
{\bf H}(x,y,z)=e^{ik_0z}\sum_nc_ne^{-iE_nz}{\bf h}_n(x,y),
}
\label{maxsol}
\end{equation}
which can serve as a basis for qualitative understanding of spectral properties of dilute systems. Such an approach allows one to pinpoint the specific details and provide information
about the energy band structure.

To close this section let us note that if the dielectric
permittivity $\epsilon (x, y)$ is a periodical function, then the spectrum
of the Hamiltonian (\ref{tem}) consists of allowed
and forbidden energy bands, if the term $\hat{V}$ is zero. If  $\hat{V}\neq 0$ then the electronic energy bands are split by $\hat{V}$ and as a consequence the polarization of the propagated light is rotated.

\subsection{Theory: TE wave}
\label{sec:capsize-theory-te}
\begin{figure}
\begin{center}
\includegraphics[width=16.0cm]{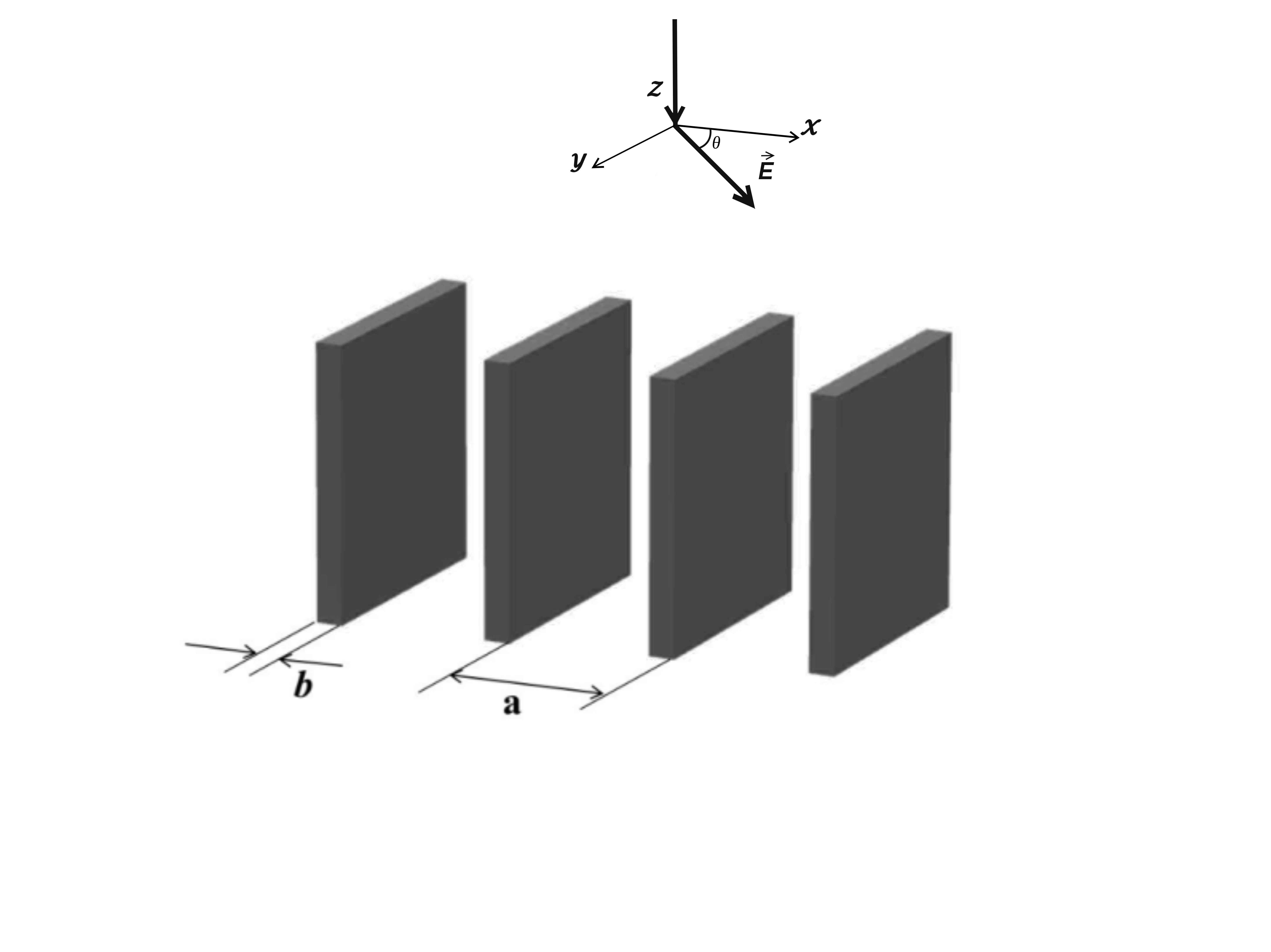}
\caption{Geometry of the problem.}
\label{pol-rot-geom}
\end{center}
\end{figure}

We considered a dielectric medium with two dimensional inhomogeneity which is expressed by
spatially dependent dielectric permittivity $\varepsilon(x,y)$. A plane wave enters the
medium from the $z < 0$ half-space at normal incidence (see Fig. \ref{pol-rot-geom}).

Using Maxwell equations one can obtain a
wave equation for E in the form
\begin{equation}
\scalemath{1}{
\nabla^2{\bf E}(x,y,z)+k_0^2\varepsilon(x,y){\bf E}(x,y,z)- \nabla (\nabla \cdot {\bf E}(x,y,z))=0,
}
\label{sqeq}
\end{equation}

where $k_0=\omega/c$ is the wave number corresponding to the angular frequency $\omega$ of an incident photon.
As we will see below, the last term in (\ref{sqeq}) causes rotation of vector $\bf E$ in the $xoy$ plain. Similarly, we seek the solution of the wave equation by representing it as a product of fast and slowly changing functions , $\bf\tilde{E}(x,y,z)$
\begin{equation}
{\bf E}(x,y,z)=e^{ik_0z} {\bf \tilde{E}}(x,y,z).
\label{slfas}
\end{equation}
Substituting Eq.(\ref{slfas}) into Eq.(\ref{sqeq}) and neglecting the second derivative of $\tilde{E}$ with respect to $z$ we get
\begin{equation}
\scalemath{1}{
i\frac{d\tilde{{\bf E}}}{dx}=\hat{H}(x,y)\tilde{{\bf E}},
}
\label{sreq1}
\end{equation}
where $\tilde{{\bf E}}\equiv\binom{\tilde{E}_x}{\tilde{E}_y}$ and $\hat{H}=\hat{H}_0+\hat{V}_{so}$. is a $2\times 2$ matrix. 
Since in a system with a small fraction of dielectric the light propagates in $z$ direction mostly forward, without back scattering we ignore the second derivative of $\tilde{E}$ with respect to z. $\hat{H}_0$ is given in the form
\begin{equation}\label{tem}
\scalemath{1}{
  \hat{H}_0=
\begin{pmatrix}
  -\frac{1}{2k_0}\nabla_t^2+\frac{k_0}{2}(1-\varepsilon(x,y)) & 0 \\
  0 & -\frac{1}{2k_0}\nabla_t^2+\frac{k_0}{2}(1-\varepsilon(x,y))
\end{pmatrix},
}
\end{equation}
where $\nabla_t^2\equiv \partial^2/\partial x^2+\partial^2/\partial y^2$ and the inhomogeneity term is given by
\begin{equation}\label{inhom}
\scalemath{1}{
 \hat{V}_{so}=\frac{1}{2k_0}
\begin{pmatrix}
\frac{\partial^2}{\partial x^2} & \frac{\partial^2 }{\partial x \partial y}\\
\frac{\partial^2 }{\partial y \partial x} & \frac{\partial^2}{\partial y^2} 
\end{pmatrix}.
}
\end{equation}
So the problem of electromagnetic wave propagation in photonic crystal is reduced to a quantum motion of a particle with mass $k_0$ in a 2d periodic potential$V(x,y)=\frac{k_0}{2}\left(1-\varepsilon(x,y)\right)$, where the spatial coordinate $z$ plays
the role of the time. Therefore, knowing the spectrum of the Hamiltonian $\hat{H}(x,y)=\hat{H}_0 +\hat{V}_{so}$ we can analyze various polarization effects in dilute photonic crystal. To do that we represent the solution of Eq. (3) through the eingenfunctions of the Hamiltonian
\begin{equation}
\scalemath{1}{
\tilde{{\bf E}}(x,y,z)=\sum_nc_ne^{-iE_nz}{\bf e}_n(x,y)
}
\label{eig}
\end{equation}
and
\begin{equation}
\scalemath{1}{
\hat{E}{\bf e}_n(x,y)=E_n{\bf e}_n(x,y).
}
\label{eigfun}
\end{equation}
Finally, substitution of Eq.(\ref{eig}) into Eq.(\ref{slfas}) yields the solution of Maxwell equation
\begin{equation}
\scalemath{1}{
{\bf E}(x,y,z)=e^{ik_0z}\sum_nc_ne^{-iE_nz}{\bf e}_n(x,y)
}
\label{maxsol}
\end{equation}

In this paragraph let us discuss the spectrum of 1d photonic crystal. Examining the problem without coupling interaction $\hat{V}_{so}\equiv 0$ in Ref. \cite{GGC16} it is shown that states at the center of the first Brilloin zone with quasimomentum $q=0$ are those which mainly contribute to the
transmitted central diffracted wave. In addition, we consider the case when the main contribution to
the central diffracted wave is given by the modes of transverse motion of a photon with negative energy
such that $E_n < k_0$ . For the symmetric Kronig-Penney model in one dimensional case dielectric function
has the form

For a one-dimensional symmetric Kronig-Penney model, discussed in this paper, the dielectric function has the form
\[
  \varepsilon(x,y) =\left\{
\begin{array}{rl}
1 & \mbox{if $0<x<a-b$} \\
\varepsilon & \mbox{if $a-b<x<a$}
\end{array} \right.
\]
Using periodicity of $u(x)$ and its derivatives, the Bloch state wave function $u(x)\equiv u_{q=0}(x)$  can be represented in the form that are valid for the energy $E_0\equiv E(q=0)$. Also those must satisfy the dispersion relation (see below, Eq. (\ref{disp})).

\begin{eqnarray}
u_{1}(x)=\frac{A}{\cosh\frac{\beta(a-b)}{2}}\cosh\beta \bigg (x-\frac{a-b}{2}\bigg),\quad 0<x<a-b \nonumber \\
u_{2}(x)=\frac{A}{\cos\frac{\alpha b}{2}}\cos\alpha \bigg (x+
\frac{b}{2}\bigg),\quad a-b<x<a
\label{wavefun}
\end{eqnarray}
where $\beta=\sqrt{2k_0E_0}$ and $\alpha=\sqrt{2k_0(V_d-E_0)}$, $E_0>0$, $V_d=\frac{k_0(\varepsilon-1)}{2}$. $A$ can be found from the normalization condition $\int_{0}^{a}dxdy u^2_{q=0}(x,y)=1$.

The Bloch state wave function $u(x,y)$  obeys the Schr\"{o}dinger equation
\begin{equation}
\scalemath{1}{
\left[-\frac{1}{2k_0}\nabla_t^2+V(x)\right]u(x)=E_0u(x).
}
\label{shrod}
\end{equation}
and energy $E_0\equiv E(q=0)$ for $q=0$ state is determined from the relation analogous to the one dimensional Kronig-Penney dispersion relation.
\begin{equation}
\scalemath{1}{
1=\cos\alpha b\cosh\beta(a-b)-\frac{\alpha^2-\beta^2}{2\alpha\beta}\sin\alpha b\sinh\beta(a-b).
}
\label{disp}
\end{equation}
Putting all of this information together one can, at least numerically, evaluate not only the energy $E_0$, using Eq. (\ref{disp}), but also calculate the intensity of central diffracted wave in two dimensional  system in terms of dielectric(metal) fraction.

One-dimensional photonic crystal case (dielectric permittivity $\varepsilon(x)$ periodically depends only one coordinate ) can be obtained from above formulas Eq.(\ref{wavefun}) and Eq.(\ref{disp}) by the following substitutions (assuming that  $y\equiv 0$): $\beta=\sqrt{2k_0E_0}$  and $\alpha=\sqrt{2k_0(V_d+E_0)}$.

After a brief sketch of the steps of calculating of spectrum of periodic photonic crystal  in one and two dimensional systems we start a detailed discussion of the interaction $\hat{V}_{sp}$ role  in a polarization rotation. We will see below that in some cases the interaction term $\hat{V}_{sp}$  plays a central role of the capsizing of initial polarization.

Now let us discuss the problem of polarization rotation in one dimensional case. In 1d case spin-orbit interaction has the form
\begin{equation}\label{spinorb1d}
\scalemath{1}{
 \hat{V}_{so1d}=\frac{1}{2k_0}
\begin{pmatrix}
\frac{\partial^2}{\partial x^2} & 0\\
0 & 0 
\end{pmatrix}.
}
\end{equation}
Interaction $\hat{V}_{o1dp}$ splits $E_0$ into two energy levels. To find energy splitting in the mean field approximation in a one dimensional case, we substitute the operator $\hat{V}_{so1d}$ by its expectation value $v$ and diagonalize $2\times 2$ Hamiltonian matrix
\begin{equation}\label{split}
\scalemath{1}{
\hat{H}=
\begin{pmatrix}
E_0+v& 0\\
0     & E_0
\end{pmatrix},
}
\end{equation}
where
\begin{equation}\label{2dspint}
\scalemath{1}{
v=\frac{1}{2k_0}\int_{0}^{a}dxu(x)\frac{\partial^2}{\partial x^2}u(x).
}
\end{equation}
Eigenvalues and corresponding eigenfunctions of Eq.(\ref{split})
\begin{eqnarray}
{\bf e}_1&=&c_1\binom{1}{0}; \quad E_1=E_0+v\nonumber\\
{\bf e}_2&=&c_2\binom{0}{1};\quad E_2=E_0,
\label{eigenvalfun}
\end{eqnarray}
where $c_{1,2}$ are arbitrary constants. Note that, we can calculate the polarization vector rotation angle when electromagnetic wave propagates through the medium. Using Eqs.(\ref{maxsol}) and (\ref{eigenvalfun}) for the electric field of central diffracted wave, we get

\begin{equation}\label{finalfield}
\scalemath{1}{
  \binom{E_x(x,y,z)}{E_y(x,y,z)}=\left [e^{i(k_0-E_1)z}{\bf e_1}+e^{i(k_0-E_2)z}{\bf e_2}\right ]u(x)
}
\end{equation}

After introducing the parameter $c=c_1/c_2$ as the ratio of $c_1$ and $c_2$ when $z=L$, for the rotation angle of the polarization vector we obtain
\begin{equation}\label{angle}
\scalemath{1}{
\tan\theta(L)=\frac{E_y(z=L)}{E_x(z=L)}=\frac{c_2}{c_1} e^{-ivL}
}
\end{equation}
given that the initial polarization angle $\theta_0$ at $L=0$ is given by
\begin{equation}\label{inangle}
\scalemath{1}{
\tan \theta_0= \frac{c_2}{c_1}.
}
\end{equation}
Using Eqs. (\ref{inangle})) and (\ref{angle}) we calculate the real part of the polarization rotation angle \begin{equation}
\label{finalangle}
\scalemath{1}{
Re[\theta(L)]=\frac{1}{2} \rm{arccot \frac{\cot2\theta_0}{|\cos v_{1d}L)|}},
}
\end{equation}
where $v_{1d}$ is the splitting parameter, which can be calculated by substituting (\eqref{wavefun}) into (\eqref{2dspint})
\begin{figure}
\begin{center}
\includegraphics[width=16.0cm]{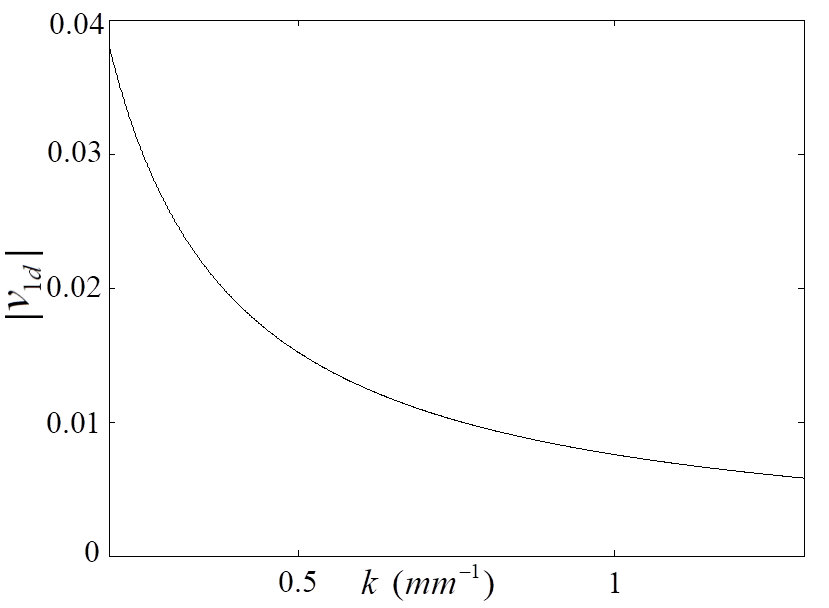}
\caption{The dependence of the splitting parameter on the wave number ($a$ = 10mm, $b$ = 0.5mm, $\epsilon$ = 10).}
\label{v1d-fig}
\end{center}
\end{figure}
\begin{equation}
\label{spl1dfin}
\scalemath{1}{
v_{1d}=\frac{A^2}{k_0} \Big[\frac{\beta^2}{\cosh^2(\frac{\beta(a-b)}{2})}\Big(\frac{\sinh(\beta(a-b))}{2\beta} + \frac{a-b}{2}\Big)-\frac{\alpha^2 \Big(\frac{b}{2} + \frac{\sin(\alpha b)}{2\alpha}\Big)}{\cos(\frac{\alpha b}{2})} \Big],
}
\end{equation}
where
\begin{equation}
\scalemath{1}{
A^{2}=\Big[\frac{1}{\cosh^2(\frac{\beta(a-b)}{2})}\Big(\frac{\sinh(\beta(a-b))}{2\beta} + \frac{a-b}{2}\Big)+\frac{ \Big(\frac{b}{2} + \frac{\sin(\alpha b)}{2\alpha}\Big)}{\cos(\frac{\alpha b}{2})} \Big]
}
\end{equation}
The dependency of the absolute value of the splitting parameter 1d v and the wavenumber k is
represented in Fig. \ref{v1d-fig}.

From Eq. (\ref{finalangle}) we see that when
\begin{equation}
\label{v1d-periodicity}
\scalemath{1}{
v_{1d}L=\frac{\pi}{2} + \pi n, \quad\quad n = 0, \pm 1 ...,
}
\end{equation}
there is abrupt change of polarization. In this case $Re[\theta(L)]$, when $0 < \theta < \pi/4$ and
$Re[\theta(L)] = \pi/2$ when $\pi/4 < \theta < \pi/2$. Thus depending on the initial polarization, for the resonant
frequencies - $\cos v_{1d}(k_0)L = 0$, $Re[\theta(L)]$ obtains only two values - 0 and $\pi/2 $. In order to find theoretical resonant frequencies we calculated the splitting parameter $v_{1d}(k_0)$ for different wavelengths
and found out that the first resonant frequency corresponds to wavenumber 1
$k_0 ~ 0.47mm^{-1}$ , which is
very close to the frequency observed experimentally in \cite{GGC16}. So, the theoretical results for resonance polarization are consistent with experimental observations.

\subsection{Spectrum of photonic crystal}
\label{sec:pc-spectrum}

Before discussing the physics induced by $\hat{V}$ coupling let us consider $\hat{V}=0$. For this case the quantum-mechanical problem
is reduced to the well-known Kronig-Penney (KP) model due to the fact that the periodic $\varepsilon(x,y)$ spectrum of Hamiltonian consists of energy bands separated by gaps and wave functions described by Bloch states.
In this section we will investigate band structure for symmetric KP model in one and two dimensional systems for
quasimomentum $q=0$, that is at the center of Brillouin zone. Note that only these states give contribution to the
transmitted light intensity in the normal direction\cite{GGC16}.

We first discuss a two-dimensional symmetric KP model where the dielectric function has the form
\begin{eqnarray}
  \varepsilon(x,y) =\left\{
\begin{array}{rl}
1 & \mbox{if $0<x,y<a-b$} \\
\varepsilon & \mbox{if $a-b<x$ or $a-b<y$}
\end{array} \right. \label{ep2}
\end{eqnarray}
The Bloch state wave function $u(x,y)\equiv u_{q=0}(x,y)$, using periodicity of $u(x,y)$ and its derivatives,  can be represented in the form that are valid for the energy $E_0\equiv E(q=0)$ that must satisfy the dispersion relation (see below, Eq. (\ref{disp})).

\begin{eqnarray}
u_{1}(x,y)=\frac{A_2}{\cos\frac{\beta(a-b)}{2}}\cos\beta \bigg (x+y-\frac{a-b}{2}\bigg)\quad 0<x,y<a-b \nonumber \\
u_{2}(x,y)=\frac{A_2}{\cos\frac{\alpha b}{2}}\cos\alpha \bigg (x+y+
\frac{b}{2}\bigg),\quad a-b<x\quad or \quad a-b<y
\label{wavefun}
\end{eqnarray}
where $\beta=\sqrt{k_0E_0}$ and $\alpha=\sqrt{k_0(V_d+E_0)}$, $E_0>0$, $V_d=k_0(\varepsilon-1)/2$. The parameter $A_2$ can be found from the normalization condition $\int_{0}^{a}dxdy u^2_{q=0}(x,y)=1$.
The Bloch state wave function $u(x,y) \equiv u_{q=0}(x,y)$  obeys the Schr\"{o}dinger equation
\begin{equation}
\scalemath{1}{
\left[-\frac{1}{2k_0}\nabla_\bot^2+V(x,y)\right]u(x,y)=E_0u(x,y),
}
\label{shrod}
\end{equation}
 and energy $E_0\equiv E(q=0)$ (for $q=0$ state) is determined from the similar to the one dimensional Kronig-Penney dispersion relation
\begin{equation}
\scalemath{1}{
1=\cos\alpha b\cos\beta(a-b)-\frac{\alpha^2+\beta^2}{2\alpha\beta}\sin\alpha b\sin\beta(a-b).
}
\label{disp}
\end{equation}
Putting all of this information together one can, at least numerically, evaluate not only the energy $E_0$, using equation  (\ref{disp}), but also calculate the intensity of central diffracted wave in $2d$ system in terms of dielectric (metal) fraction.

As for the $1d$ PC case (dielectric permittivity $\varepsilon(x)$ periodically depends only one coordinate) it can be obtained from above formulas by the following substitutions (assuming that $y\equiv 0$): $\beta=\sqrt{2k_0E_0}$ and $\alpha=\sqrt{2k_0(V_d+E_0)}$.
We have presented the photonic band structure of transversal motion for the $1d$ crystal used in our experiment in Fig.5.
\begin{figure}
 \begin{center}
\includegraphics[width=16.0cm]{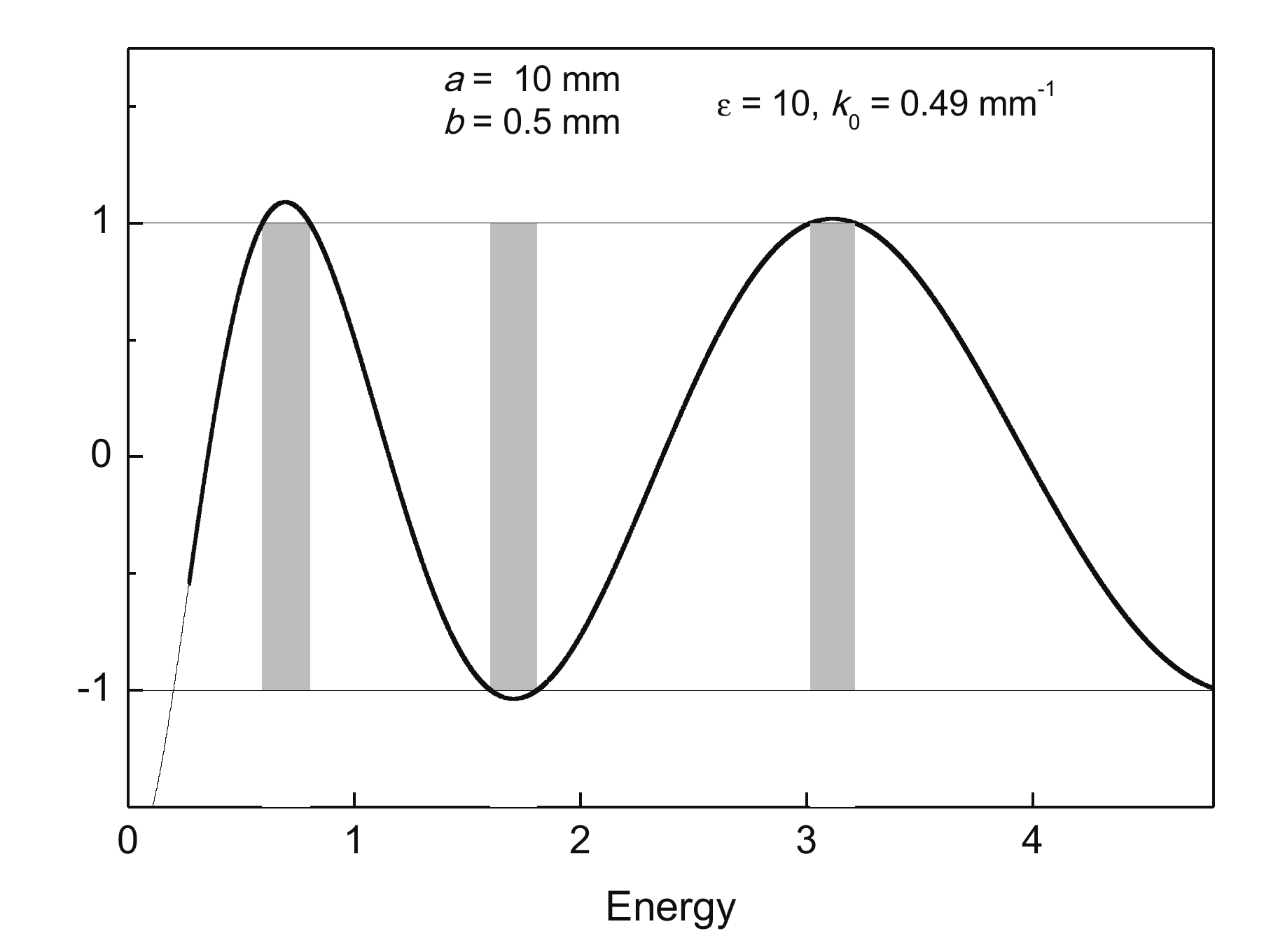}
\caption{Transversal photonic band structure. The vertical axis is the dispersion equation right side Eq.(\ref{disp}) calculated for the $1d$ crystal used  in our experiment Fig.2 at the resonance wavenumber $k_0=0.49mm^{-1}$, Fig.3. The shaded parts are the forbidden bands.}
\label{fig.5}
\end{center}
\end{figure}

The border value $E_0\equiv E(q=0)=0.401<k_0$ is the transversal wavenumber. The incident wavenumber $k_0$ lies in the forbidden band.
After a brief sketch of the steps of calculating of spectrum of periodic PC in $1d$ and $2d$ systems we start a detailed discussion of the interaction $\hat{V}$ role in a PR. We will see below that in some cases the interaction term $\hat{V}$  plays a central role of the capsizing of initial polarization.

\subsection{Polarization rotation: 1D and 2D photonic crystals}
\label{sec:polarization-rotation}
As mentioned above the interaction $\hat{V}$ splits $E_0$ into two energy levels. To find energy splitting in the
mean field approximation in a $2d$ case, we substitute the operator $\hat{V}$ by its expectation value
and diagonalize $2\times 2$ matrix Hamiltonian
\begin{equation}\label{split}
\scalemath{1}{
\hat{H}=
\begin{pmatrix}
E_0-v_{2d}& v_{2d}\\
v_{2d}   & E_0-v_{2d}
\end{pmatrix},
}
\end{equation}
where
\begin{equation}\label{2dspint}
\scalemath{1}{
v_{2d}=\frac{1}{2k_0}\int_{0}^{a}dxdy\frac{1}{\varepsilon(x,y)}\frac{\partial \varepsilon(x,y)}{\partial y}u(x,y)\frac{\partial u(x,y)}{\partial y}.
}
\end{equation}
Note that  for the symmetrical Kronig-Penney model all the integrals in $\hat{V}$ are equal to $v_{2d}$.
Direct calculation  of the integral (\ref{2dspint}) by using  (\ref{wavefun}) yields
\begin{equation}
v_{2d}=A_2^2\frac{\varepsilon-1}{2k_0\varepsilon}\tan{\frac{\alpha b}{2}}{\sin2\alpha b}
\label{split2d},
\end{equation}
where $A_2$, as was mentioned, is the normalization parameter with dimensionality $a^{-1}$ (a is the period of the system).

For the magnetic field of central diffracted wave we get
\begin{equation}\label{finalfield}
\scalemath{1}{
  \binom{H_x(x,y,z)}{H_y(x,y,z)}=\left [e^{i(k_0-E_1)z}{\bf h_1}+e^{i(k_0-E_2)z}{\bf h_2}\right ]u(x,y).
  }
\end{equation}

One can interpret this result as an PR due to the interaction term $\hat{V}$ or Rashba splitting parameter $v$. The latter,  imposing asymmetry into a periodic 2d structure in $xy-$plane, leads to a different speed in the x and y axes resulting in a phase delay between $H_x$ and $H_y$ components.

Introducing parameter $c=c_1/c_2$, the PR angle
at $z=L$ can be written in the form
\begin{equation}\label{angle}
\scalemath{1}{
\tan\theta(L)=\frac{H_y(z=L)}{H_x(z=L)}=\frac{c^2-1-2ic\sin (E_1-E_2)L}{c^2+2c\cos (E_1-E_2)L+1},
}
\end{equation}
provided that an initial polarization angle $\theta_0$ at $L=0$ is given by
\begin{equation}\label{inangle}
\scalemath{1}{
\tan \theta_0= \frac{c-1}{c+1}.
}
\end{equation}
By applying equation  (\ref{inangle}) and calculating the real part of rotation angle $\theta$ from equation  (\ref{angle})
one obtain ($E_1-E_2\equiv 2v_{2d}$)

\begin{equation}\label{finalangle}
\scalemath{1}{
{\rm Re}\, \theta(L)=\frac{1}{2} {\rm arccot} (\cot2\theta_0\cos2v_{2d}L),
}
\end{equation}
where the principal interval of the arccotangent function is chosen $0\le {\rm arccot}\, x <\pi$ to
avoid discontinuity at $x = 0$.

The above expression ${\rm Re}\, \theta(L)$ is a general expression for PR angle and valid for  dilute photonic crystal with inhomogeneity in the plane perpendicular to the propagation direction. The value of ${\rm Re}\, \theta(L)$ strongly depends on the frequency of the electromagnetic wave, traveling distance $L$ and on the initial polarization angle $\theta_0$.

Some obvious results that follow from equation  (\ref{finalangle}) above. It is clear, that for the particular values $v_{2d}L=\pi n$ ($n=0, 1, \dotsc$) the ${\rm Re}\, \theta(L)=\theta_0$.
In the limit $2v_{2d}L<<1$ the ${\rm Re}\, \theta(L) \approx \theta_0$ and completely independent of a particular value of $\theta_0$.
Next, if the initial angle $\theta_0 \to \pi/4$  ( $c\gg 1$- maximum difference between two orthogonally polarized eigenmodes of polarization) the ${\rm Re}\, \theta(L) \to \pi/4$ for all $L$. This means that the initial wave polarization at $\pi/4$ does not change when crossing  the medium and becomes independent of the traveling distance.

However, a significant rotation of ${\rm Re}\, \theta(L)$ from the initial value $\theta_0$ occurs when $\cos 2vL$ being positive changes the sign. In fact, if $\cot{2\theta_0} \to \infty$, then the $Re\theta(L)$ asymptotically tends to zero, if $\cos 2v_{2d}L>0$ and tends to $\pi/2$ if $\cos 2v_{2d}L<0$. Hence, for the particular value of $\theta_0=0$ (or $c=1$), the jump from $0$ to $\pi/2$ of the ${\rm Re}\, \theta(L)$ occurs at $v_{2d}L=\pi/4$. This quite drastic change of the initial polarization can be completely controlled having the appropriate parameters $L$ and frequency-dependent $v$ (see Fig. 2). In a special case of $\cos 2v_{2d}L=-1$ the $\theta(L)=\pi/2-\theta(L=0)$ and we recover the result of Ref. \cite{dielpolrot12}.

Note that capsize effect takes place for all values of $2vL$ in the range $[\pi/2, 3\pi/2]$ including the value $2vL=\pi$, considered in Ref. \cite{dielpolrot12}. When $vL=\pi/4$, as we mentioned, ${\rm Re}\, \theta(L)=\pi/4$ and independent of the initial polarization, beside $\theta_0=0$.

As another manifestation of the drastic change of the polarization, one can study the behavior of the imaginary portion of the $\theta(L=0)$ (the ellipticity or the ratio of ellipse axes) at resonant frequencies $\cos 2v_{2d}(k_0)L=0$. The latter can be written in the compact form, based on Eqs. (\ref{angle}) and (\ref{inangle})
\begin{equation}\label{im2d}
\scalemath{1}{
{\rm Im}\, \theta(L)  =\frac{1}{2}\ln|\tan\theta_0|.
}
\end{equation}
This expression provides useful information about the ratio of ellipse axes (within the sign accuracy) near the particular angle $\theta_0 \approx 0$. Exactly at $\theta_0=0$ the imaginary portion, ${\rm Im}\, \theta(L)$, tends to infinity. This means that light remains linearly polarized after traveling distance $L$ while the polarization direction rotates by $\pi/2$.

Knowing the explicit form of the inhomogeneity term (\ref{inhom}) it is not
difficult to calculate the splitting interaction $\hat{V}$ in the $1d$ case
\begin{equation}\label{spinorb1d}
\scalemath{1}{
  \hat{V}_{1d}=\frac{1}{2k_0\varepsilon}
\begin{pmatrix}
  0 & 0\\
  0 & -\frac{d\varepsilon}{d x}\frac{d}{d x}
\end{pmatrix}.
}
\end{equation}
The  splitting parameter $  v_{1d}$ reads
\begin{equation}
\scalemath{1}{
  v_{1d}=\frac{1}{2k_0}\int_{0}^{a}dx \frac{u(x)}{\varepsilon(x)}\frac{d\varepsilon}{d x}\frac{d u(x)}{d x},
}
\end{equation}
where $\varepsilon(x)=1+(\varepsilon-1)\Theta(x-(a-b))$ is the $1d$ periodic dielectric constant in a unit cell $[0,a]$.
Using (\ref{wavefun}) and evaluating this integral, we find
\begin{equation}\label{spl1d}
\scalemath{1}{
v_{1d}=\frac{A_{1d}^2\alpha(\varepsilon-1)}{2k_0\varepsilon}\tan\frac{\alpha b}{2},
}
\end{equation}
where $A_{1d}$ is the normalization parameter with dimensionality $a^{-1/2}$. Comparing splitting parameters in 2d, equation (\ref{split2d}), and 1d, equation (\ref{spl1d}), one can see that the 2d splitting parameter consists of an additional $b/a\ll 1$ multiplier. As a consequence, the small $v_{2d}$ in dilute photonic crystals leads to a large optical path difference between the two polarizations and make difficult to observe experimentally a capsize in 2d dilute crystals. This statement was supported by our numerical calculations of the splitting parameters $v_{1d}$ and  $v_{2d}$ versus $k_0$ using the structural parameters of the photonic crystal, used in experimental purposes. Indeed, as seen from Fig. 6, for $b=0.5$ mm, $a=10$ mm, and $\varepsilon=10$ there is a large factor in between $v_{1d}$ and $v_{2d}$ parameters. As we will see below, due to the small value of $v_{2d}$ in the frequencies range of our set up, we were not able clearly see the capsize in 2d case. However, in 1d case the capsize takes place precisely at the resonance frequencies, predicted by theory.

In a similar fashion, demonstrated in the previous section, while calculating the real part of rotation angle $Re\theta(L)$ (see equation  \ref{finalangle}), one can find PR angle for $1d$ case
\begin{eqnarray}
\label{rotangle1d}
  {\rm Re}\, \theta(L) &=& \frac{1}{2} {\rm arccot}\frac{\cot 2\theta_0}{|\cos v_{1d}L|}.
 \end{eqnarray}
The only difference between the two equations (\ref{finalangle}) and (\ref{rotangle1d}) (without taking into account the factor 2) is that the ${\cos v_{1d}L}$ in 1d case appears in the denominator. A direct consequence of this flip is that the drastic change of the polarization now takes place at $v_{1d}L= \pi/2+\pi n, n=0,\pm 1,..$. In this case ${\rm Re}\, \theta(L)=0$ if $0<\theta_0<\pi/4$ and ${\rm Re}\, \theta(L)=\pi/2$ if $\pi/4<\theta_0<\pi/2$. So for the resonant frequencies $\cos v_{1d}(k_0)L=0$, ${\rm Re}\, \theta(L)$ acquires only two values $0$ and $\pi/2$ depending on initial polarization angle $\theta_0$. Capsize takes place exactly at $\theta_0=\pi/4$.

As for the imaginary part of rotation angle that describes the ellipticity of transmitted wave at resonance frequencies $\cos v_{1d}(k_0)L=0$ one can write
\begin{equation}
\label{im1d}
\scalemath{1}{
{\rm Im}\, \theta(L)=\frac{1}{2}\ln\sqrt{2}|\tan(\pi/4-\theta_0)|.
}
\end{equation}
The infinity value of imaginary part at $\theta_0=\pi/4$ is an clear evidence of a linear polarization and a change of an initial polarization {\bf to} $\pi/2$.

\begin{figure}
\begin{center}
\includegraphics[width=16.0cm]{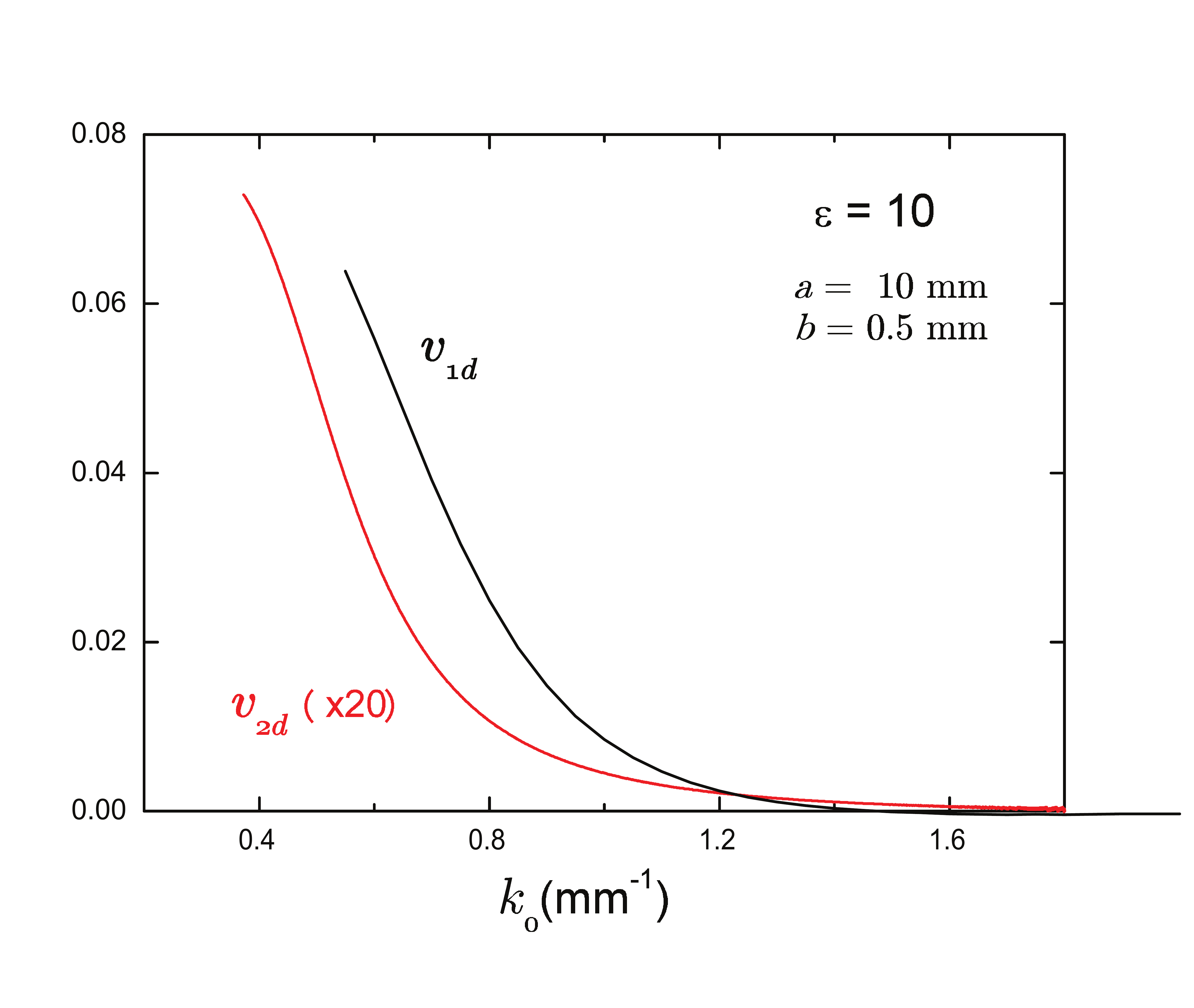}
\caption{Splitting parameter. The wavenumber dependence of the splitting parameter $v$ for one and two
dimensional crystals. In order to make a direct comparison $v_{2d}$ has been multiplied by a factor 20.}
\label{fig.6}
\end{center}
\end{figure}

As mentioned above the interaction $\hat{V}$ splits $E_0$ into two energy levels. To find energy splitting in the
mean field approximation in a $2d$ case, we substitute the operator $\hat{V}$ by its expectation value
and diagonalize $2\times 2$ matrix Hamiltonian
\begin{equation}\label{split}
\scalemath{1}{
\hat{H}=
\begin{pmatrix}
E_0-v_{2d}& v_{2d}\\
v_{2d}   & E_0-v_{2d}
\end{pmatrix},
}
\end{equation}
where
\begin{equation}\label{2dspint}
\scalemath{1}{
v_{2d}=\frac{1}{2k_0}\int_{0}^{a}dxdy\frac{1}{\varepsilon(x,y)}\frac{\partial \varepsilon(x,y)}{\partial y}u(x,y)\frac{\partial u(x,y)}{\partial y}.
}
\end{equation}
Note that  for the symmetrical Kronig-Penney model all the integrals in $\hat{V}$ are equal to $v_{2d}$.
Direct calculation  of the integral (\ref{2dspint}) by using  (\ref{wavefun}) yields
\begin{equation}
v_{2d}=A_2^2\frac{\varepsilon-1}{2k_0\varepsilon}\tan{\frac{\alpha b}{2}}{\sin2\alpha b}
\label{split2d},
\end{equation}
where $A_2$, as was mentioned, is the normalization parameter with dimensionality $a^{-1}$ (a is the period of the system).

\subsection{Concluding remarks}
\label{sec:capsize-conclusion}
In this chapter we have investiagated the resonance polarization change in photonic crystal. Firstly, we present the expermental set-up that was used to detect the polarization change. In the following two sections it is presented the theory which is based on Maxwell's equations with two dimensional inhomogeneous permittivity. In \cite{gevdav19} we have proposed the theory for the TE waves and obtained a reasonable correspondence between the theoretical and experimental resonant frequencies.

\section[Extended symmetries in geometrical optics]
		{Extended symmetries in geometrical optics}
\label{sec:extended-symmetries}

\subsection{Introduction}
\label{sec:extended-symmetries-intro}
We examine additional symmetries of specific refraction index profiles that are used in the well-known phenomena of perfect imaging and cloaking. In the considered cases, the translation generator and the angular momentum are conserved. We express the ray trajectory parameters through the integrals of motion and observe the existence of a photon state with maximal angular momentum, which can be used as an optical resonator. Application in plasmons and the role of polarization are discussed, and the spin Hall effect in an extended symmetry profile is predicted.

We consider the refraction indices which yield  the optical metrics  coinciding with those of three-dimensional sphere (for positive $\kappa$) and two-sheet hyperboloid/ pseudosphere (for negative $\kappa$),
 \begin{equation}
n({\bf r})=\frac{n_0}{|1+\kappa{\bf r}^2|},\qquad \kappa=\pm\frac{1}{4r^2_0}.
\label{ns}
\end{equation}
While the generic homogeneous spaces have $so(3)$ symmetry algebra generated by conserved angular momentum, three-dimensional sphere and pseudosphere have symmetry algebras  $so(4)$ and $so(3.1)$, respectively. The reason for that is the existence of three additional conserved quantities - ``translation generators".
The two-dimensional counterpart of such refraction index with a positive sign along with other profiles corresponding to the two-dimensional Hamiltonian systems with closed trajectories \cite{Leonhardt06}was already used to describe cloaking phenomena in $2d$ via conformal mapping. This profile is well-known  in optics  as  ``Maxwells fish eye" \cite{Maxwell},\cite{bowolf} and was  debated as a possible tool for perfect imaging \cite{Leonhardt09,Philbin10,Blaikie10}.

In our opinion, the importance of the present consideration is the direct use of extended symmetry of the sphere and pseudosphere. Seemingly, it might be extended to other three-dimensional profiles corresponding to the Hamiltonian systems with close trajectories, such as three-dimensional oscillator and Coulomb potential and their deformations with Calogero-like potentials, as well as their generalizations to three-dimensional spheres and pseudospheres  \cite{CalCoul}. The consideration of the three-dimensional (two-sheet) hyperboloid which can be used for the study of plasmon perfect imaging, cloaking is another important point in the given study. Here we have mentioned cloaking and perfect imaging as possible applications of closed ray trajectories. However, there can be other applications too.

In the coming sections we mostly neglect the light polarization, whose interaction with the inhomogeneity of dielectric permittivity leads to many well-known effects, e.g. the optical Hall effect \cite{Mur04, bnature, bliokh}, capsize of polarization and straightening of light in dilute photonic crystals \cite{ghgc17,ggc19}, etc. We only briefly discuss its influence on the trajectories postponing the detailed consideration for the future study.

\subsection{Initial relations}
\label{sec:extended-symmetries-initial-relations}

\begin{figure}
 \begin{center}
\includegraphics[width=16cm]{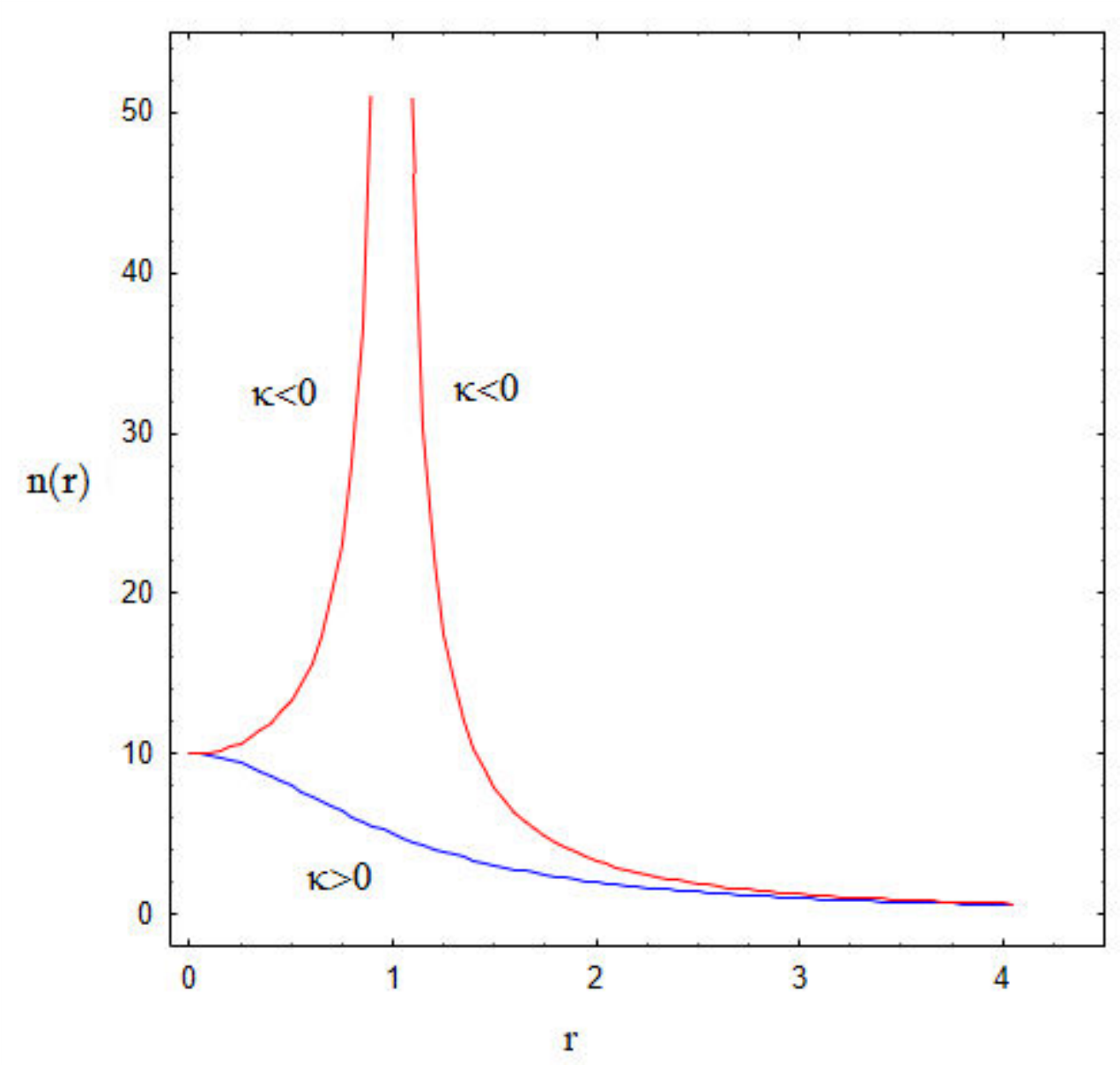}
\caption{Refraction index profiles for both positive and negative values of $\kappa$.}
\label{mfe-profiles-for-different-kappa}
\end{center}
\end{figure}

It is well-known  that minimal action principle came \textit{to} physics from the
geometric optics. Initially, it was invented for the description of
the propagation of light in media by Fermat, and is presently known as the Fermat principle,
\begin{equation}
 {\cal S}_{Fermat} =\frac{1}{\lambdabar_{0}}\int d{s},\qquad ds=n({\bf r}) |d{\bf r}/d\tau |d\tau
\label{gactions2}\end{equation}
where  $n({\bf r})$ is the
refraction index, and $\lambdabar_0$ is wavelength in vacuum which defines the length of the light trajectory under the
assumption that the helicity of light is neglected. This  action could be interpreted as the action of the system on the
three-dimensional curved space equipped with the ``optical metrics" (or Fermat metrics)
of Euclidean signature
 \begin{equation}
  ds^2= g_{AB} dx^A dx^B,\qquad
g_{AB}=n^2({\bf r})\delta_{AB}, \qquad A,B=1,2,3
\label{om}
\end{equation}
In the special case when the
refraction index has a form \eqref{ns},
the Fermat metrics  describes  three-dimensional sphere (for  $\kappa>0$), or two-sheet hyperboloid (for  $\kappa<0$) \textit{with} radius $r_0$.
In this case, in addition to rotational,  $so(3)$, symmetry  which yields the conserving angular momentum, the system has $so(4)$/$so(3.1)$ symmetry which provides the  three-dimensional sphere/hyperboloid, by  three additional symmetries conserved quantities .
In the case of positive $\kappa$, the refraction index is a decreasing function of $r$, while for the negative $\kappa$, it has an increasing part (see Fig. \ref{mfe-profiles-for-different-kappa}).
In this paper, we show that due to the extended symmetry mentioned above, these very different profiles lead to closed ray trajectories.
As noted in the Introduction,  the case of the positive sign was already used to describe cloaking phenomena in $2d$ via conformal mapping \cite{Leonhardt06}.

Let us give the Hamiltonian formulation of the system defined by the action \eqref{gactions2}.
Due to its reparametrization-invariance, the Hamiltonian constructed by the standard Legendre transformation is identically zero. However, following Dirac's theory \cite{dirac}, the constraint between  momenta  and coordinates appears as follows
 \begin{equation}
  \Phi\equiv g^{ij}(x)p_ip_j-\lambdabar^{-2}_0 =0, \quad{\rm with}\quad g^{ij}p_ip_j=\frac{{\bf p}^2}{n^2({\bf r})},\qquad {\rm with}\quad g^{ij}g_{jk}=\delta^i_k.
 \label{constraint}
 \end{equation}
 Hence, the Hamiltonian system corresponding to the action \eqref{gactions2},
is defined by the canonical Poisson brackets
 \begin{equation}
  \{f,g\}=\frac{\partial f}{\partial x_\alpha}\frac{\partial g}{\partial p_\alpha}-\frac{\partial f}{\partial p_\alpha}\frac{\partial g}{\partial x_\alpha}
 \label{pb0}\end{equation}
and by the  Hamiltonian
\begin{equation}
\mathcal{H}_0=\alpha({\bf p},{\bf r})\Phi =\alpha({\bf p},{\bf r})\left( \frac{{\bf p}^2}{n^2({\bf r})}-\lambdabar^{-2}_0 \right)\approx 0.
\label{h0}
\end{equation}
Here $\alpha$ is the Lagrangian multiplier, which could be an arbitrary function of coordinates and momenta.
When we write down the Hamiltonian equations of motion, the notation ``weak zero" ($\mathcal{H}_{0}\approx 0$) indicates that we should take into account the constraint \eqref{constraint} only after differentiation,
\be
{\dot f}({\bf r}, {\bf p})=\{f, \mathcal{H}_0\}=\{f,\alpha\}\Phi+\alpha \{f,\Phi\}\approx\alpha\{f,\Phi\}.\ee
The  arbitrariness in the choice of the function $\alpha$ reflects the reparametrization-invariance of the action \eqref{gactions2}.
Suppose, for the description of the  equations of motion in terms of arc-length of the original Euclidian space  one should choose (see \cite{bliokh})
\begin{equation}
\alpha^{-1} =\frac{|{\bf p}|+ \lambdabar^{-1}_0 n({\bf r})}{n^2({\bf r})},\qquad \Rightarrow\quad \mathcal{H}_0=|{\bf p}|- \lambdabar^{-1}_0n({\bf r})
\label{alpha}
\end{equation}
With this choice, the Hamiltonian equations of motion take the conventional form \cite{ko}
 \begin{equation}
{\bf \dot{p}}=\lambdabar^{-1}_0{\bf \nabla} n(\bf r),\qquad
{\bf \dot{r}}={\bf p}/{|{\bf p}|}.
\label{hameq}
\end{equation}
These equations describe the motion of a wave package with center coordinate ${\bf r}$ and momentum ${\bf p}$ in a curved space.
However, for preserving the similarity with classical mechanics we will deal  with the generic formulation \eqref{h0}, i.e. we will not fix the parametrization of light rays.

We are interested in the integrals of motion, i.e. physical quantities that are conserved along the ray trajectory.
For the isotropic media, when  the refraction index is a spherical symmetric function, $n({\bf r})=n(|{\bf r}|)$, the angular momentum of the system is conserved
\begin{equation}
\mathbf{L}=\mathbf{r}\times \mathbf{p}.
\label{moment}
\end{equation}
When the refraction index \textit{has} the form \eqref{ns},  the Fermat   metrics coincide with those of three-dimensional sphere (for $\kappa >0$) or two-sheet hyperboloid (for $\kappa <0$) written in conformal flat coordinates (we ignore here $n_0$ factor),
 \begin{equation}
\frac{d{\bf r}\cdot d{\bf r}}{(1\pm \frac{{\bf r}^2}{4r^2_0})^2}=\left(d{\bf y} \cdot d{\bf y} \pm dy^2_4\right)\mid_{ y^2_4\pm {\bf y}^2=r^2_0} \; ,
 \end{equation}
where
 \begin{equation}
 y_4=-r_0\frac{1-\kappa {\bf r}^2}{1+\kappa {\bf r}^2},\quad {\bf y}=\frac{{\bf r}}{1+\kappa {\bf r}^2},\quad {\rm with }\quad \kappa =\pm\frac{1}{4r^2_0}
 \label{amc}
 \end{equation}
 From Eq.\eqref{amc} we get
 \begin{equation}
 {\bf r}=2r_0\frac{{\bf y}}{r_0+y_4}\quad \Rightarrow\quad {\bf r}^2=4r^2_0\frac{r_0-y_4}{r_0+y_4}
 \label{rest}
 \end{equation}
This is just stereographic projection of the (pseudo)sphere on the ``$\mathbf{r}$-space"  which touches it at the pole $y_4=-r_0$.
The upper and lower hemispheres are projected to the inside and outside of the three-dimensional ball with radius $2r_0$, respectively. The "equatorial sphere" $y_4=0$  is projected to the boundary of that ball,  which is a two-dimensional sphere in the ``$\mathbf{r}$-space". Due to  the $so(4)/so(3.1)$ symmetry of three-dimensional sphere/hyperboloid, in addition to  $so(3)$ algebra generators \eqref{moment},  the system  possesses  three more  conserving quantities
\begin{equation}
\mathbf{T}=(1-\kappa \mathbf{r}^2)\mathbf{p}+2\kappa(\mathbf{pr})\mathbf{r}\quad :\quad \{\mathbf{T},\mathcal{H}_0\}=0.
\label{T0}
\end{equation}
The Hamiltonian becomes  Casimir of   $so(4)/so(3.1)$ algebra(s)
\begin{equation}
\frac{{\bf p}^2}{(1+\kappa r^2)^2}={\bf T}^2+4\kappa{\bf L}^2\quad\Rightarrow \quad  {\bf T}^2+4\kappa{\bf L}^2=\frac{n^2_0}{\lambdabar^{2}_0}.
\label{cas}
\end{equation}
Notice also, that $\mathbf{T}$ is perpendicular to $\mathbf{L}$:  $\mathbf{T}\bot \mathbf{L}$.

\subsection{Trajectories}
\label{sec:extended-symmetries-trajectories}

\begin{figure}
   \begin{center}
\includegraphics[width=16cm]{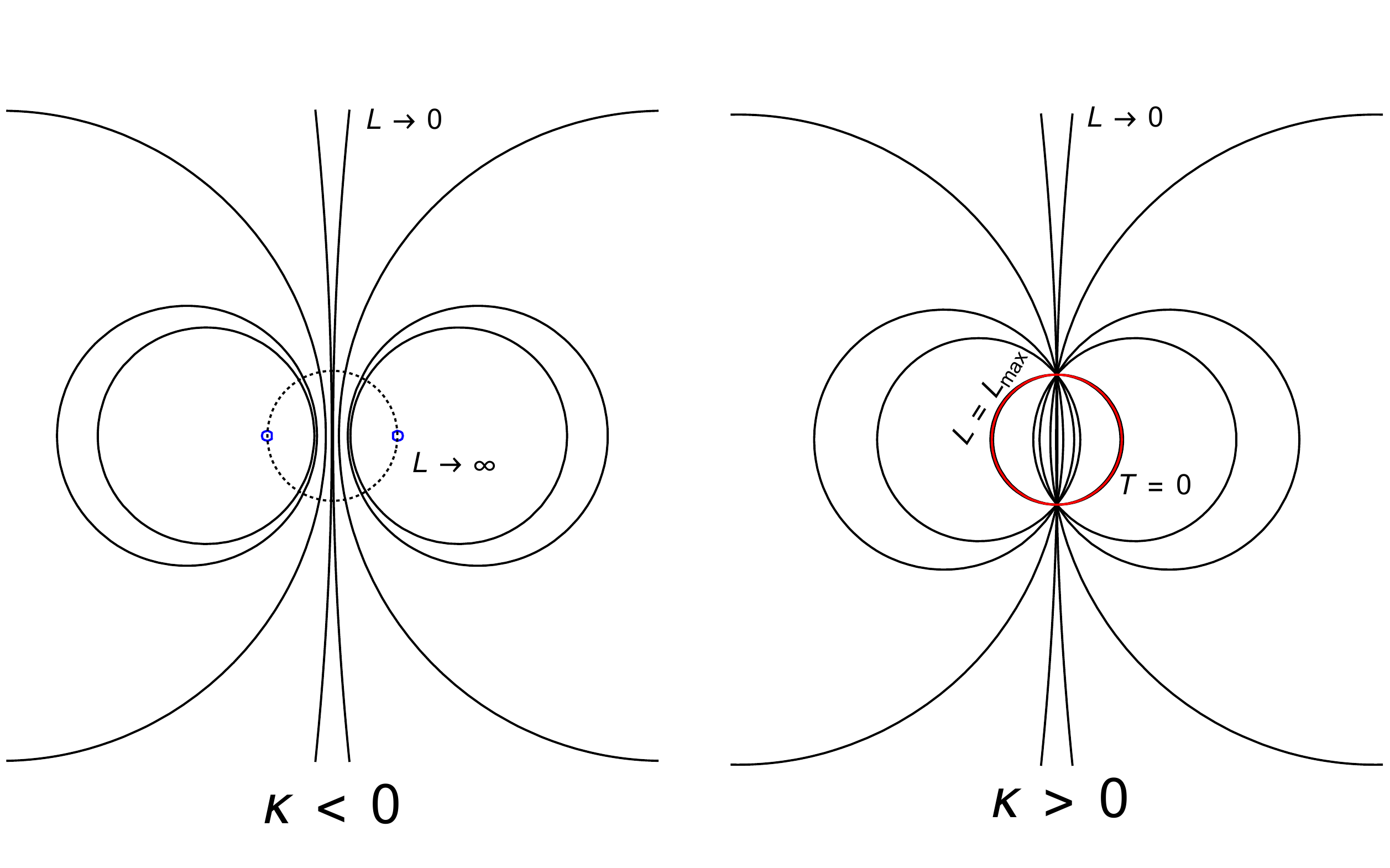}
\caption{Ray trajectories for different values of ${\bf T},{\bf L},\kappa$. Dashed and red circles with radius $2r_0$ are given by the refraction profile \eqref{ns}. The red circle is the photon trajectory with maximal angular momentum of the photon in $\kappa>0$ case. In $\kappa>0$ case, symmetric trajectories with $\pm a_0$ intersect while in $\kappa<0$ case they do not. The straight lines are trajectories without angular momentum $L=0$. The small blue circles are trajectories in $L\to\infty$ limit  provided that $\kappa<0$.}
\label{photon-trajectories-for-different-kappa}
\end{center}
\end{figure}

Extended symmetry allows to obtain ray trajectories without solving the equations of motion \eqref{hameq}. Namely, the vector product of   ${\bf T}$ and  ${\bf r}$  immediately yields  the expression of trajectories,
\begin{equation}
\mathbf{L}=\frac{\mathbf{T}\times\mathbf{r}}{1-\kappa \mathbf{r}^2}\quad \Rightarrow\quad     \vert\mathbf{r}-\mathbf{a}_0\vert^2=R^2_0,\qquad \mathbf{a}_0\equiv\frac{\mathbf{T}\times{\mathbf L}}{2\kappa L^2}, \quad R_0=\frac{n_0}{2|\kappa| L\lambdabar_0}.
\label{trajec}
\end{equation}
Ray trajectories are the circles with center ${\bf a_0}$ and radius $R_0$ \cite{bowolf}. Note  that  the radius of  circle is independent of the sign of $\kappa$, whereas the coordinates of the center ${\bf a_0}$ depend on the sign of $\kappa$. From \eqref{trajec}, it is easy to find that
  \begin{equation}
  |a_0|=\sqrt{R_0^2-1/\kappa }
  \label{centcor}
  \end{equation}
Using expressions \eqref{centcor} and \eqref{trajec} one can
  draw the ray trajectories Fig. \ref{photon-trajectories-for-different-kappa}.
For the photons with zero angular momentum $L=0$,  the radius of the circle goes to infinity $R_0\to\infty$, i.e. we get a straight line.
In contrast with common approaches (see, e.g.  \cite{bowolf}), we express the equations of ray characteristics  \eqref{trajec} through the integrals of motion, which allows us to consider different physical situations.

Now, let  us consider the  cases when either ${\bf T}$ or ${\bf L}$ become  zero (they  cannot  be  equal to zero simultaneously due to \eqref{cas}).
 If $\kappa<0$ from Eq. \eqref{cas} it follows that the only possibility is $L=0$ , $T\neq 0$, with $T$ taking minimal value
 $T_{min}=n_0/\lambdabar_0$.
Using  \eqref{centcor},\eqref{cas} one can see that  for $\kappa<0$  one always has $|a_0|>2r_0$ and for $\kappa>0$ one has  $R_0>2r_0$. So, in both cases there are not any ray closed trajectories inside the  area of radius $2r_0$, see Fig. \ref{photon-trajectories-for-different-kappa}. In $\kappa>0$ case, photons with angular momentum $L>L_{max}$ will not form closed trajectories and correspondingly can not ensure perfect imaging and cloaking. This restriction is absent in $\kappa<0$ case. In the conformal mapping scheme cloaking area is determined by the outer part of closed trajectories \cite{Leonhardt06}. Therefore from \eqref{trajec}, it follows that cloaking area will disappear for small $L$.

For $\kappa >0$ , $T=0$, $L\neq 0$ angular mommentum  $L$ acquires the maximal  value $L_{max}=n_0r_0/\lambdabar_0$ \eqref{cas}.
As it follows from \eqref{T0},   when $T=0$, $r^2=4r_0^2$ and ${\bf p}\bot {\bf r}$ the  photon trajectory becomes a circle with the radius $R_0=2r_0$ and with center at $\mathbf{a}_0=0$ (see Fig.\ref{photon-trajectories-for-different-kappa}).
In this state, the angular momentum of photon  can be very large $n_0r_0/\lambdabar_0\gg 1$. These states are interesting for quantum information purposes \cite{zeil18}. Besides that, this state can be used as an optical resonator that is an accumulator of energy.
To be convinced let us determine electric and magnetic fields on this trajectory. In geometrical optics one can use the following expansions \cite{bowolf}
\begin{equation}
{\bf E}=e^{\imath \frac{\Psi}{\lambdabar_0}}\sum_{m\geq 0}(-\imath \lambdabar_0)^m{\bf e}_m\qquad
{\bf H}=e^{\imath \frac{\Psi}{\lambdabar_0}}\sum_{m\geq 0}(-\imath \lambdabar_0)^m {\bf h}_m,
 \label{exp}
 \end{equation}
 where  ${\bf e_m,h_m}$ are functions of coordinates which  can be found by substituting expressions of $\bf E$ and $\bf H$ from \eqref{exp} into Maxwell equations. Geometrical optics approximation corresponds to leading terms of expansion \eqref{exp}. The equation for eikonal $\Psi$ has the form
\begin{equation}
n({\bf r})\frac{d {\bf r}}{ds} ={\bf \nabla}\Psi
\label{eikeq}
\end{equation}
Substituting \eqref{hameq} into \eqref{eikeq} and using \eqref{h0} we then  find
\begin{equation}
\Psi=\lambdabar_0{\bf pr}
\label{eik}
\end{equation}
As mentioned above, for this trajectory $T=0$, $L=L_{max}=n_0r_0k_0$, ${\bf p}\bot {\bf r}$ and ${\bf p}{\bf r}\equiv 0$. Therefore on this trajectory, as it follows from \eqref{exp}, the phase (eikonal) remains constant during the round trip of the photon. Hence, in this state, photon constructively interferes with itself, and the energy is being accumulated (see also \cite{svelto10,turks14}).

Note that the basic ray trajectory ($T=0$, $L=L_{max}$) with large angular momentum is similar to whispering gallery modes that originate as an eigenstate of dielectric sphere (see \cite{oraevsky02}).

\subsection{Plasmon}
\label{sec:extended-symmetries-plasmon}

In the $\kappa<0$ case,  the refraction index diverges at the point  $r = 2r_0$. Such a situation can be realized,
for example, on the metal surfaces near the plasmon resonance frequencies.
Indeed, it is  well known that  dispersion equation of plasmon on the interface of a metal with dielectric constant $\varepsilon(\omega)<0$
and with dielectric permittivity $\varepsilon_d$ has the form
\begin{equation}
k_p=\frac{\omega}{c}\sqrt{\frac{\varepsilon(\omega)}{\varepsilon(\omega)+\varepsilon_d}},
\label{plasmon}
\end{equation}
with $\varepsilon(\omega)\approx -\varepsilon_d$ near the plasmon resonance.

Suppose that the dielectric material has an inhomogeneous profile $\varepsilon_d\equiv \varepsilon_d(r)$.  From the expression above, it can be presumed that
a plasmon moves in a $2d$ medium with refraction profile
\begin{equation}
n(r)=\sqrt{\frac{\varepsilon(\omega)}{\varepsilon(\omega)+\varepsilon_d(r)}}.
\label{refract}
\end{equation}
Suppose that  $\mathbf{r}_p$ is a resonance point,  $\varepsilon(\omega)= -\varepsilon_d(r_p)$.
Expanding $\varepsilon_d(r)$ around $r_p$ and assuming that $\varepsilon'(r_p)=0$, we get
\begin{equation}
n(r)\approx \frac{1}{|r-r_p|}\sqrt{\frac{2\varepsilon(\omega)}{\varepsilon_d''(r_p)}}.
\label{appref}
\end{equation}
If we choose $n_0$ and $r_0$ such that
\begin{equation}
n_0r_0=\sqrt{\frac{2\varepsilon(\omega)}{\varepsilon_d''(r_p)}}, \qquad \kappa<0,\quad r_0=\frac{r_p}{2},
\label{compref}
\end{equation}
near  the  resonance  point, $n(r)$ will obtain the form \eqref{ns}.
So, it is possible to choose indexes $\varepsilon_p (r)$ such that
near plasmon resonance point, closed trajectories and therefore cloaking phenomenon via conformal mapping can be realized (see also \cite{plasmon}).

\subsection{Generalizations}
\label{sec:extended-symmetries-generalizations}
In the previous section we  related closed trajectories  with the free particles  on the sphere and two-sheet hyperboloid, which are the simplest three-dimensional maximally superintegrable systems (the $N$-dimensional dynamical system is called maximally superintegrable when it has $2N-1$ functionally independent integrals of motion. In these  systems all trajectories are closed).
It was argued in \cite{Leonhardt06}, that the cloaking phenomenon takes place when  all trajectories of the dynamical system  defining the
refraction index are closed. In other words, dynamical system should be maximally superintegrable.
However, only the oscillator and the Coulomb problem on Euclidian spaces were considered in the mentioned paper.  At the same time,
there are their well-known generalizations to the spheres and two-sheet hyperboloids  defined by the potentials
\cite{higgs}
\be
 V_{osc}=\frac{\omega^2{\bf r}^2}{(1-\kappa{\bf r}^2)^2},\qquad V_{Coul}=-\gamma \frac{|1-\kappa {\bf r}|}{r},
\ee
as well as their further superintegrable deformations  including, in particular, the Calogero-like term \cite{CalCoul}.

Considering the energy surface $\mathcal{H}-E =0$ as a constraint \eqref{constraint}   and properly rescaling  the potential $V({\bf r})$ along with value of energy $E$, one gets
the modified profiles which can be used  for describing cloaking and perfect imaging phenomena
\begin{equation}
\mathcal{H}={(1+\kappa r^2)^2}{{\bf p}^2}+V(r), \quad \Rightarrow \quad \widetilde{n}({\bf r})=n_0\frac{\sqrt{|1-V({\bf r})|}}{|1+\kappa{\bf r}^2|}.
\end{equation}
The addition of Calogero-like term breaks spherical symmetry of the profile at the same time preserving its superintegrability.
However, in this case symmetry algebra is highly nonlinear which may cause troubles in the description of closed ray trajectories in a purely algebraic way.

Another way to find the profiles which should possess perfect imaging is to perform the simple canonical transformation $({\bf p}, {\bf r})\to (-{\bf r}, {\bf p})$. In this case the energy surface
takes a form $|{\bf r}|=n(p)$. Then expressing $p$ via $r$, we will get the new profile admitting cloaking, given by the function $\widetilde{n}_{inv}$ which is inverse to \textit{the} initial profile $n(r)$:
 $\widetilde{n}_{inv}\left( n(r)\right)=r$.
For example, it transforms the initial profile \eqref{ns}  to the one associated with the Coulomb problem
\be
(1+\kappa p^2)^2r^2=\frac{n^2_0}{\lambdabar^2_0}\quad\Rightarrow\quad   {n}_{inv}=\sqrt{\frac1{\kappa} \left(\frac{n_0}{\lambdabar_0 r}-1\right)},
\label{nCoul}\ee

\subsection{Inclusion of polarization}
\label{sec:extended-symmetries-polarization-included}
\begin{figure}
\begin{center}
\includegraphics[width=16cm]{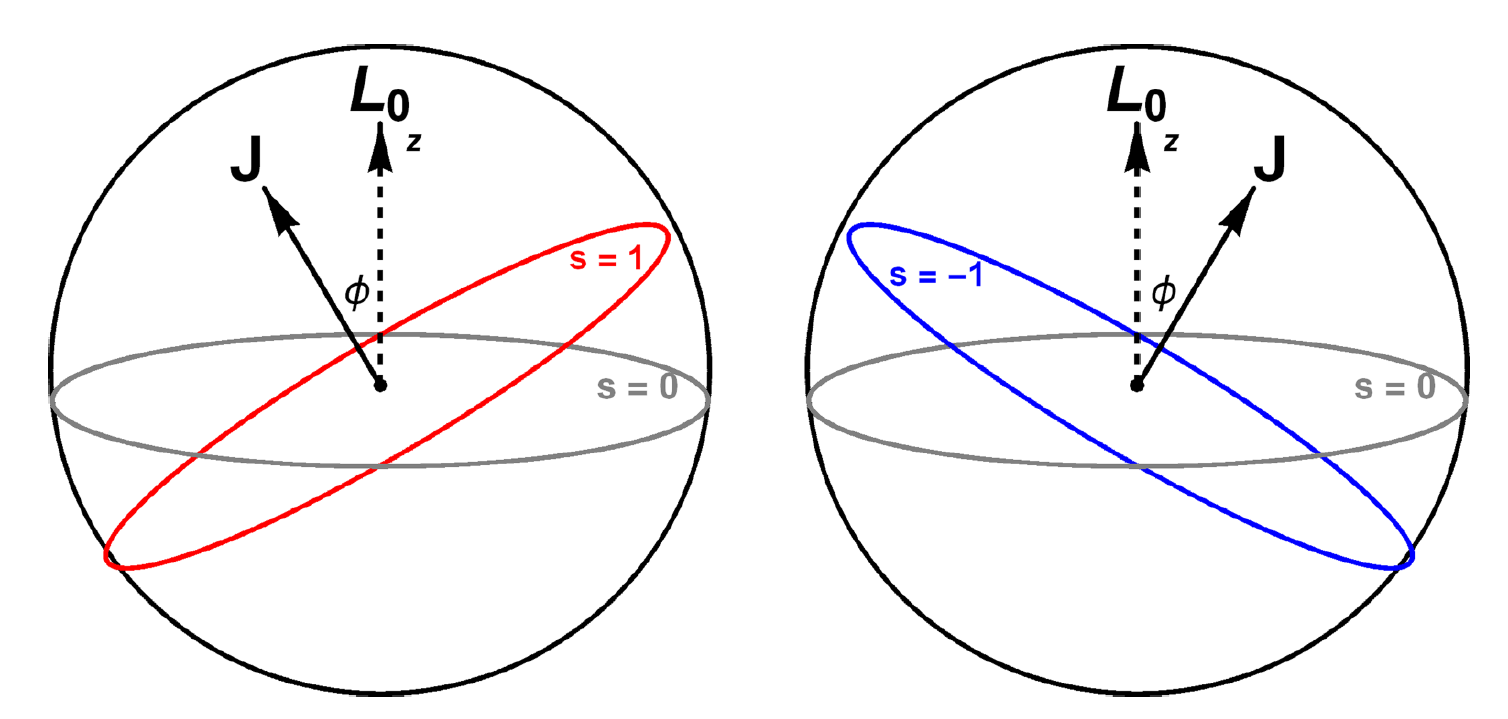}
\caption{Basic ray trajectories for photon with different polarizations. $\phi$ is the trajectory plane rotation angle.}
\label{trajectory-rotations}
\end{center}
\end{figure}

Let us briefly discuss the inclusion of polarization.
To  this end we should add to the Lagrangian \textit{the term} ${\bf p\dot r}$, the vector-potential  of ``Berry monopole" $\mathbf{ A({p})\dot p}$ i.e. by the potential of the Dirac monopole located at the origin of \textit{the} momentum space \cite{bliokh}
\begin{equation}
\frac{\partial}{\partial \mathbf{p}}\times \mathbf{A}(p)=\frac{\mathbf{p}}{|\mathbf{p}|^3}
\label{monop}
\end{equation}
From the  viewpoint of Hamiltonian formalism this means that
 we should preserve the form of \textit{the} Hamiltonian \eqref{h0} and replace the initial Poisson brackets \eqref{pb0} by the modified ones
 \begin{equation}
  \{f,g\}=\frac{\partial f}{\partial x_\alpha}\frac{\partial g}{\partial p_\alpha}-\frac{\partial f}{\partial p_\alpha}\frac{\partial g}{\partial x_\alpha}-\frac{sp_k}{p^3}\varepsilon_{klm}\frac{\partial f}{\partial x_l}\frac{\partial g}{\partial x_m}
 \label{pbm}
 \end{equation}
  where $s$ is the spin of the photon, which is equal to one for circularly polarized photon and to zero  for linearly  polarized photon.
  The above deformation of Poisson bracket violates  the  symmetry of the Hamiltonian, and therefore, can break the closed trajectories. However the basic trajectory ($T=0, L=L_{max}$) in the limit $s\to 0$ preserves its form (see below).
  Using new definition of \textit{the} Poisson bracket Eq.(\ref{pbm}), one gets the equations of motion in the form

 \begin{eqnarray}
\dot{\bf r}=\frac{{\bf p}}{p}+\frac{s {\bf L}}{\lambdabar r p^3}\frac{\partial n}{\partial r},\quad \dot{\bf p}=\lambdabar^{-1}{\bf \nabla}n(r)\quad
\dot{\bf T}=\frac{2s\kappa {\bf L}}{\lambdabar r p^3}\frac{\partial n}{\partial r}({\bf pr}) \nonumber\\
 \dot{\bf L}=\frac{s}{\lambdabar r}\frac{\partial n}{\partial r}\left(\frac{{\bf p}({\bf pr})}{p^3}-\frac{{\bf r}}{p}\right),\quad
\dot{\bf S}=-\frac{s}{\lambdabar r}\frac{\partial n}{\partial r}\left(\frac{{\bf p}({\bf pr})}{p^3}-\frac{{\bf r}}{p}\right)
\label{eqmodexp}
\end{eqnarray}
where spin vector is determined as ${\bf S}=s{\bf p}/p$. When spin variable is taken into account in the spherical symmetrical refraction index profile $n(r)$, from Eq.(\ref{eqmodexp}) it follows that the total angular momentum ${\bf J}={\bf L}+{\bf S}$ is preserved, $\dot{\bf J}\equiv 0$. We will develop perturbation theory on $s$ exploring Eqs.(\ref{eqmodexp}). In the first-order perturbation theory on $s$, one can substitute all the terms containing $s$ by their zero-order values (values when $s=0$).
Using this approach, we can see that, in the first-order approximation, the value of $T$ on the basic trajectory is zero: $T(s)=0$.
Scalarly  multiplying \textit{$\bf J$ by $\bf r$}, for the basic trajectory, one gets ${\bf Jr}=0$. In this approximation, it follows from Eq.(\ref{trajec}) that $a_0=0,\quad R=2r_0$ and $|{\bf r}|^2=4r_0^2$. This means that basic trajectory remains a circle \textit{from} the same sphere with the center at the origin. However the plane of the circle is rotated and now is perpendicular to ${\bf J}$ and not to
${\bf L_0}$ as in $s=0$ case. The rotation angle can be found by scalarly multiplying ${\bf J}$ and ${\bf p}$: ${\bf Jp}=sp$ and therefore $\cos\theta=s/J$ and $\sin\phi=s/J$, where $\theta$ and $\phi\approx\pi/2-\theta$ are angles between $\bf J$ and $\bf p$ and $\bf J$ and $\bf L_0$, respectively (see Fig. \ref{trajectory-rotations}).

In the first order on $s$, $J$ can be substituted by $L_{max}=n_0r_0/\lambdabar$, \quad $sin\phi=s\lambdabar/n_0r_0\ll 1$.  Hence the actual perturbation parameter is $s\lambdabar/n_0r_0\ll 1$ therefore perturbation theory can be applied for $ s=\pm 1$ as well. The sign of rotation angle depends on the sign of $s$. So for right hand circular polarized and left hand circular polarized photons one will have different trajectories on the sphere. This is an analogue of spin Hall effect \cite{Mur04, bnature, bliokh}  in Maxwell fish eye refraction profile.



\subsection{Concluding remarks}
\label{sec:extended-symmetries-conclusion}
In this chapter we have examined the additional symmetries of Maxwell Fish eye and have shown that the ray trajectories in Maxwell fish eye are closed and the expressions for ray trajectories are found directly from the integrals of motion. Moreover, we have shown that there exists a photon state with maximal angular momentum and it can be used as a possible means for creating optical resonators. Light polarization was mostly neglected in this section. However, a perturbation theory has been proposed which illustrates that when we take into account the spin the trajectories rotate by a fixed angle. However, light polarization breaks the the hidden symmetries of the system, so one needs more sophisticated approach in order to preserve the symmetries of the system. That is what we do in the next chapter.

\section[Maxwell fish eye  for polarized light]
		{\textbf{Maxwell fish eye  for polarized light}}
\label{sec:fish-eye-polarized}

\subsection{Introduction}
\label{sec:fish-eye-polarized-intro}

It is well-known that the minimal action principle came in physics from geometric optics. Initially, it was invented for the description of the propagation of light and is presently known as the Fermat principle
\begin{equation}
 {\cal S}_{Fermat} =\frac{1}{\lambdabar_{0}}\int d{\tilde l},\qquad d{\tilde l}:=n(\mathbf{ r}) |d\mathbf{r}/d\tau |d\tau
\label{gactions2}
\end{equation}
where  $n(\mathbf{ r})$ is the
refraction index, and $\lambdabar_0$ is the wavelength in vacuum. This  action
could be interpreted as the action of the system on the
three-dimensional curved space equipped with the ``optical metrics" or the Fermat metrics (see. \cite{arnold})
of Euclidean signature
 \begin{equation}
  d{\tilde l}^2= n^2(\mathbf{r})d\mathbf{r} \cdot d\mathbf{r}\;.
\label{om}
\end{equation}
Thus, the symmetries of the system which describe the propagation of light in a particular medium are coming from the symmetries of the optical metrics of that particular medium. In systems with a maximal number of functionally independent integrals of motion ($2N-1$ integrals for $N$-dimensional system), all the trajectories of the system become closed.
The closeness of the trajectories makes such profiles highly relevant in the study of cloaking and perfect imaging phenomena.
The most well-known profile of this sort is the so-called ``Maxwell fish eye" profile which is defined by the metrics of (three-dimensional) sphere or pseudosphere (under pseudosphere we mean the upper (or lower) sheet of the two-sheet hyperboloid).
\begin{equation}
n_{Mfe}(\mathbf{r}) =\frac{n_0}{|1 +  \kappa  \mathbf{r}^2 |}, \qquad \kappa=\pm\frac{1}{4r^2_0}
\label{Mfe0}
\end{equation}
where the plus/minus sign in the expression for $\kappa$ corresponds to the sphere/pseudosphere with the radius  $r_0$, and $n_0>0$.
Apart from applications in cloaking and perfect imaging phenomena \cite{Pendry06,Leonhardt06,leonhardconformal}, Maxwell fish eye is a common profile in quantum optics with single atoms and photons \cite{perczel}, optical resonators \cite{turks14}, discrete spectrum radiation \cite{discretespectrumradiation} etc.
Moreover,  there are many experimental implementations of the  Maxwell fish eye lenses \cite{zhao2021bifunctional,lu2020millimeter,koala2020}.

In most of the listed studies, scalar wave approximation was used and polarization of electromagnetic waves was not taken into account. In these systems, the ray trajectories belong to the plane which is orthogonal to the angular momentum.  Introduction of spin (polarization) results to the rotation of this plane by a constant angle proportional to spin, moreover,  it breaks the non-rotational symmetries of the optical systems with Maxwell fish eye profile, so that photon trajectories no longer remain closed   \cite{gevdav20}. Thus, the key property of the Maxwell fish-eye profile which makes it relevant in cloaking and perfect imaging phenomena is violated.

In the present paper, we continue our study of a polarized light passing through the Maxwell fish eye profile within the geometrical optics approximation. The key point of our study is that we  introduce a   polarization-dependent  deformation of the Maxwell fish-eye profile
\be
n^{s}_{Mfe}(\mathbf{r})= \frac{n_{Mfe}(\mathbf{r})}{2} \left(1 + \sqrt{1 - \frac{4\kappa s^2\lambdabar^2_0 }{n_0}\frac{1}{n_{Mfe}(\mathbf{r})}}\right),
\label{spinFishEye0}
\ee
where $n_{Mfe}(\mathbf{r})$ is the original Maxwell fish eye profile given by \eqref{Mfe0}, and $s$ is the light polarization. For the linearly and circularly  polarized light we have $s=0$ and $s=1$ respectively.
The proposed deformation restores all the symmetries of the optical Hamiltonian, with Maxwell fish eye profile, which were broken after the inclusion of polarization. It also ensures the closeness of the trajectories for polarized photons and can be used for cloaking and perfect imaging of polarized photons.
It is seen, that spin induced term is proportional to the dimensionless parameter $s^2\lambda_0^2/r_0^2$ where $r_0$ is the characteristic length of the profile defined in \eqref{Mfe0}. This means that spin will play a significant role only in the vicinity of wave and geometrical optics border $s\lambda_0/r_0\sim 1$, since  we are working in the framework of geometrical optics approximation, $\lambda_0\ll r_0$. Below we will investigate the influence of spin (polarization) on the ray trajectories  in the deformed  Maxwell fish eye profile given by \eqref{spinFishEye0}.\\


\subsection{Scalar waves}
\label{sec:fish-eye-polarized-scalar-waves}

Due to reparametrization-invariance of the action   \eqref{gactions2},  the Hamiltonian constructed by the standard Legendre transformation is identically zero. However, the constraint between  momenta  and coordinates appears there
\begin{equation}
\Phi:= \frac{{\bf p}^2}{n^2({\bf r})} -\lambdabar^{-2}_0 =0.
\label{constraint}
\end{equation}
Hence, in accordance with the Dirac's constraint theory \cite{dirac}
the respective  Hamiltonian system
is defined by the
canonical Poisson brackets
\begin{equation}
\{x_i, p_j\}=\delta_{ij},\quad   \{p_i, p_j\}=\{x_i, x_j\} =0,
\label{pb0}\end{equation}
and by the  Hamiltonian
\begin{equation}
\mathcal{H}_0=\alpha({\bf p},{\bf r})\Phi =\alpha({\bf p},{\bf r})\left( \frac{{  p}^2}{n^2({\bf r})}-\lambdabar^{-2}_0 \right)\approx 0.
\label{h0}
\end{equation}
Here $\alpha$ is the Lagrangian  multiplier  which could be an arbitrary function of coordinates and momenta, and $i,j=1,2,3$.
The notation ``weak zero", $\mathcal{H}_{0}\approx 0$,  means that when writing down the Hamiltonian equations of motion,
we should take into account the constraint \eqref{constraint} only after the differentiation,
\be
\frac{df({\bf r},{\bf p})}{d\tau} =\{f, \mathcal{H}_0\}=\{f,\alpha\}\Phi+\alpha \{f,\Phi\}\approx\alpha\{f,\Phi\}.\ee
The  arbitrariness in the choice of the function $\alpha$ reflects the reparametrization-invariance of \eqref{gactions2}.
For the description of the  equations of motion in terms of arc-length of the original Euclidian space one should choose (see,  e.g.  \cite{bliokh})
\begin{equation}
\alpha =\frac{n^2({\bf r})}{ {  p} + \lambdabar^{-1}_0 n({\bf r})},\qquad \Rightarrow\quad \mathcal{H}_{\rm Opt}= p - \lambdabar^{-1}_0n({\bf r}).
\label{alpha}
\end{equation}
With this choice, the  equations of motion take the conventional form \cite{ko}
\begin{equation}
\frac{d{\bf {p}}}{dl}=\lambdabar^{-1}_0{\bf \nabla} n({\bf r}),\qquad
\frac{d{\bf {r}}}{dl}=\frac{\bf p}{ {  p} },
\label{hameq}
\end{equation}
where $dl:=\alpha({\bf r}, {\bf p})d\tau$ is the element of arc-length.
These equations describe the motion of a wave package with center coordinate ${\bf r}$ and momentum ${\bf p}$ in the medium with refraction index $n({\bf r})$.\\

Assume we have a Hamiltonian system given by the Poisson bracket \eqref{pb0} and by the Hamiltonian
\be
H=\frac{{  p}^2}{2g({\bf r})}+V({\bf r}).
\label{h}\ee
In accordance with the Mopertuit principle, after fixing the energy surface $H=E$, we can relate its trajectories with the optical Hamiltonian \eqref{h0} with the refraction index
\be
 n( {\bf r})=\lambdabar_0\sqrt{2g({\bf r})(E-V({\bf r}))}.
\label{rih} \ee
Clearly, the optical Hamiltonian \eqref{h0} (as well as the Hamiltonian  \eqref{alpha}) with the  refraction index  \eqref{rih} inherits all the symmetries and constants of motion of the Hamiltonian \eqref{h}.

Canonical transformations preserve the symmetries of the Hamiltonians and their level surfaces. Hence, we are able to construct the physically non-equivalent optical Hamiltonians (and refraction indices) with the identical symmetry algebra.
The simplest illustration is the well-known relation between the Coulomb Hamiltonian which defines the so-called Coulomb refraction index profile
and the free-particle Hamiltonian on the three-dimensional sphere, which defines the ``Maxwell fish eye" refraction index (see e.g. \cite{perelomov}).
Firstly, we fix the energy surface of the  Coulomb Hamiltonian and get the respective refraction index
\begin{equation}
H_{Coul}-E :=\frac{{ p}^2}{2}-\frac{{\gamma }}{  r }-E=0, \quad \Rightarrow \quad n_{Coul}=\lambdabar_0\sqrt{2(E+\gamma/r ) },\qquad {\rm where}\quad \gamma>0.
\label{Coulsur}\end{equation}

The constants of motion of the Coulomb problem (and of the respective optical Hamiltonian) are given by the rotational momentum and by the Runge-Lenz vector
\be
\mathbf{L}=\mathbf{r}\times \mathbf{p},\qquad {\bf A} = {\bf L }\times\mathbf{p} + \gamma  \frac{{\bf r}}{  {r} }
\label{A}\ee
which form the algebra
\be
 \{A_i,A_j\}=-2\varepsilon_{ijk}H_{Coul}L_k,\quad \{A_i,L_j\}=\varepsilon_{ijk}A_k,\quad \{L_i,L_j\}=\varepsilon_{ijk}L_k.
 \label{Coulalg}\ee
Now, let us  perform a simple canonical transformation,
\be
({\bf p}, {\bf r})\to (- {\bf r} , {\bf p}).
 \label{ct}\ee
As a result, the first equation in  \eqref{Coulsur} reads
 \be
  {r}^2-\frac{2{\gamma }}{  p }-2E=0\quad\Rightarrow\quad   p  -\frac{2\gamma}{{r}^2-2E}=0
 \ee
 Interpreting the second equation as an optical Hamiltonian, we get the   refraction index profile known as the ``Maxwell fish eye" \eqref{Mfe0}  with the parameters $\kappa$ and $n_0$ defined as follows
  \be
 \kappa:= -\frac{1}{2E} ,\qquad \frac{n_0}{\lambdabar_0}:=  2\epsilon\kappa\gamma,
 \label{gammaE}\ee
 where $\epsilon=-{\rm sgn}(r^2+1/\kappa)$.

The integrals of motion \eqref{A} result in the symmetry generators of the optical Hamiltonian with the Maxwell fish eye refraction index
\be
\mathbf{L}\to \mathbf{L},\quad \mathbf{A}\to \frac{\mathbf{T}}{2\kappa},\quad \mathbf{T}=\left(1- \kappa  r^2\right)\mathbf{p} + 2 \kappa  (\mathbf{rp}) \mathbf{r}=\left(2-\frac{n_0}{n_{Mfe}({\bf r})}\right)\mathbf{p}+ 2 \kappa  (\mathbf{rp}) \mathbf{r}.
\ee
These integrals form  the $so(4)$ algebra for $\kappa >0$, and $so(1.3)$ algebra for $\kappa<0$:
\be
\{L_i,L_j\}=\varepsilon_{ijk}L_k,\quad \{T_i,L_j\}=\varepsilon_{ijk}T_k,\quad \{T_i,T_j\}=4\kappa \varepsilon_{ijk}L_k\; .
\label{Mfealg}\ee

In the  next sections  we will use the above  described duality for the  construction of the Maxwell fish eye profile for polarized light.

\subsection{Inclusion of polarization}
\label{sec:fish-eye-polarized-spin-inclusion}
Let us briefly discuss the inclusion of polarization.

To  this end we should add to the  scalar Lagrangian $L_0={\bf p\dot{r}}-p+\lambda_0^{-1}n$ the additional term $L_1=-s\mathbf{A}( \mathbf{p})\dot{\mathbf{p}}$,
where $s$ is the spin of the photon, and $\mathbf{ A}$ is the
the vector-potential  of the ``Berry monopole"  (i.e. the
  potential of the magnetic (Dirac) monopole located at the origin of momentum space) \cite{bliokh}
  \be
\mathbf{F}:=\frac{\partial}{\partial \mathbf{ p}}\times \mathbf{A}(\mathbf{p})=\frac{\mathbf{p}}{  {p} ^3}
\label{F}\ee

From the  Hamiltonian viewpoint this means to preserve the form of the Hamiltonian \eqref{h0}
and  replace the  canonical Poisson brackets \eqref{pb0}
by the twisted  ones
 \begin{equation}
  \{x_i, p_j\}=\delta_{ij},\qquad \{x_i, x_j\}=s\varepsilon_{ijk}F_k(\mathbf{p}), \qquad\{p_i, p_j\} =0,
 \label{pbB}\end{equation}
  where $ i,j,k=1,2,3$, and  $F_k$ are  the components of the Berry monopole \eqref{F}.
On this phase space the rotation generators take  the form
  \be
\mathbf{J}=\mathbf{r}\times \mathbf{p}+ s\frac{{\mathbf p}}{ {p} }
\label{Js}
\ee
while
the equations of motion read
\be
\frac{d\mathbf{p}}{dl}=\lambdabar_0^{-1}\mathbf{\nabla} n(\mathbf{r}),\qquad \frac{d\mathbf{r}}{dl}=\frac{\mathbf{p}}{ p } - \frac{s}{\lambdabar_0}\mathbf{F} \times \mathbf{\nabla}n({\bf r}),
  \ee
However, the above procedure, i.e. twisting the Poisson bracket with preservation of the Hamiltonian,  violates  the non-kinematical (hidden) symmetry of the system.
To get the profiles admitting the symmetries in the presence of polarization, we use the following observation \cite{lnp} (see \cite{mardoyan} for its quantum counterpart). Assume we have the three-dimensional rotationally-invariant system
 \be
 \mathcal{H}_0=\frac{{p}^2}{2g(r)}+V(r),\qquad  \{x_i, p_j\}=\delta_{ij}, \quad  \{p_i, p_j\}=\{x_i, x_j\}=0.
 \ee
For the inclusion of  interaction with magnetic monopole, we should  transit from the canonical Poisson brackets to the twisted ones:
\be
\{x_i, p_j\}=\delta_{ij}, \qquad  \{p_i, p_j\}=s\varepsilon_{ijk}\frac{x_k}{r^3},\qquad \{x_i, x_j\}=0.
\label{PBsp}\ee
The rotation generators  then read
\be
\mathbf{J}=\mathbf{r}\times \mathbf{p}+ s\frac{{\mathbf r}}{r}\;: \quad \{J_i,J_j\}=\varepsilon_{ijk}J_k.
\label{Jr}\ee
By modifying the initial Hamiltonian to
\be
\mathcal{H}_s=\frac{{p}^2}{2g( r)}+\frac{s^2}{2g(r){r}^2}+V(r),
\label{Hs}\ee
we find that trajectories of the system preserve their form, but the plane which they belong to, fails to be orthogonal to the
the axis $  {\mathbf{J}} $. Instead, it  turns to the constant angle
\be
\cos\theta_0=\frac{s}{|\mathbf J|}.
\ee
For the systems with hidden symmetries, one can find the appropriate modifications of the hidden symmetry generators respecting the inclusion of the monopole field.

To apply this observation on the systems with polarized light,   we should choose the appropriate integrable system with magnetic monopole, and then perform the canonical transformation
\eqref{ct}
which yields the Poisson brackets for polarized light \eqref{PBsp}. Afterwards we need to solve the following equation
 \be
  r^2+\frac{s^2}{p^2}-2 g(p) (E-V(p))=0,\quad\Rightarrow\quad p=\frac{n^s_{inv} ( r)}{\lambdabar_0}.
 \ee
For example, to get the ``polarized Coulomb profile"  we have  to start  from the free-particle Hamiltonian on three-dimensional  sphere/hiperboloid interacting with Dirac monopole: 
\be
H_{s}= \frac{(1+\kappa r^2)^2}{2}\left({p}^2 +\frac{s^2}{r^2}\right).
\ee
Then, after fixing the energy surface $H_{s}=E$ and performing canonical transformation \eqref{ct}
we arrive to the third-order (with respect to $ {p}^2$) algebraic equation:
   \be
(1+\kappa p^2)^2\left(r^2+\frac{s^2}{p^2}\right)=2E (>0),\qquad{\Leftrightarrow}\qquad  y^3u-y^2(u-\kappa s^2)-Ey+E=0,
\label{nCoul1}\ee
with $ y:=1+\kappa p^2$, $u:=r^2$.

This equation has either one real and two complex solutions or three real solutions, which  describe the ``polarized Coulomb profiles".

Conversely, when we start from the Coulomb problem with Dirac monopole we will arrive to the ``polarized Maxwell fish eye", i.e. the deformation of the ``Maxwell fish eye"  which preserves,  in the presence of polarized light, all symmetries of initial scalar system.
The latter is considered in detail in the next section.

\subsection{Polarized Maxwell fish eye}
\label{sec:fish-eye-polarized-deduction}
Let us consider  the Coulomb system   with Dirac monopole which is known as ``MICZ-Kepler system" \cite{MICZ}.
It is defined by the twisted Poisson brackets \eqref{PBsp} and  by the Hamiltonian
\begin{equation}
H_{MICZ} =\frac{{  p}^2}{2}+\frac{s^2}{2r^2}-\frac{\gamma}{r}.
\end{equation}
Besides the conserved angular momentum  \eqref{Jr},
this system has the conserved Runge-Lenz vector
\begin{equation}
\mathbf{A}_s=\mathbf{J}\times \mathbf{p}+\gamma \frac{\mathbf{r}}{r},
\label{rls}\end{equation}
which forms the symmetry algebra of Coulomb problem \eqref{Mfealg} (with the replacement $(\mathbf{L},\mathbf{A})\to (\mathbf{J},\mathbf{A}_s)$).
After performing  canonical transformation \eqref{ct}, we get
\be
H_{MICZ}=E\quad\Leftrightarrow \quad r^2+\frac{s^2}{p^2}-\frac{2\gamma}{p}-2E=0.
\ee
Solving this quadratic equation for $p$,  we get the refraction index
given by the expression \eqref{spinFishEye0},
where the notation \eqref{gammaE} is used.
%

The rotation generator \eqref{Jr} transforms to \eqref{Js}, and  the Runge-Lenz vector \eqref{rls} transforms to ${\mathbf{T}_s}/{\kappa}$, where

\be
\mathbf{T}_s= \Big(2-\frac{n_0}{n^s_{Mfe}(\mathbf{r})}\Big)
\mathbf{p}+2\kappa (\mathbf{rp})\mathbf{r} +\frac{2\kappa s}{n^s_{Mfe}(\mathbf{r})} \mathbf{J} .
\ee
Along with \eqref{Js}, these generators form the symmetry algebra of the original Maxwell fish eye profile \eqref{Mfealg} (where the pair $(\mathbf{L},\mathbf{T})$ is replaced by  $(\mathbf{J},\mathbf{T}_s)$).
The Casimirs of the symmetry algebra are given by the expressions
\be
\mathbf{T}^2_s +4\kappa(\mathbf{J}^2-s^2)=\frac{n^2_0}{\lambdabar^2_0}, \qquad \mathbf{T}_s\cdot\mathbf{J}=\frac{s n_0}{\lambdabar_0}.
\ee
Hence, for $\kappa>0$ the vectors  $\sqrt{4\kappa}\mathbf{J}$ and $\mathbf{T}_s$ form the parallelogram with the fixed lengths of  diagonals
\be
 |\mathbf{T}_s\pm \sqrt{4\kappa}\mathbf{J}|=|\frac{n_0}{\lambdabar_0}\pm \sqrt{4\kappa}s|.
\ee
This immediately leads to the conclusion that for $\kappa >0$ the   generators $\mathbf{T}_s $ and $\mathbf{J}$ reach the lower/upper bounds being parallel to each other\be
 \left( |\mathbf{J} |_{\rm min}=s,\; |\mathbf{T}_s|_{\rm max}= \frac{n_0}{\lambdabar_0}\right),\qquad \left(  |\mathbf{J} |_{\rm max}=\frac{n_0}{\lambdabar_0\sqrt{4\kappa}}\;, |\mathbf{T}_s|_{\rm min}= \sqrt{4\kappa}s \right) .
\ee
Notice also, that for   $\kappa > 0$ we get  a  restriction of rays in the finite domain
\be
\kappa > 0\;:\quad r\leq \sqrt{\frac{n^2_0}{4s^2\lambdabar^2_0\kappa^2 }-\frac1\kappa}.
\ee
One can also  note that spin appears in the expression for  the refraction index \eqref{spinFishEye0} along with the factor $\kappa {\lambda_0}^2 = ({\lambda_0}/{2r_0})^2$. In order to stay within the bounds of geometrical optics approximation, this factor must be reasonably small.
Therefore, the influence of the spin will be far more notable within certain range of distance from the core of the fish eye. The latter happens when the condition ${4\kappa s^2\lambdabar^2_0 }/{n_0} \approx n_{Mfe}(\mathbf{r})$ holds.
At these distances the refraction index in presence of spin can be much  smaller as compared to the refraction index with zero spin.
\begin{figure}
\begin{center}
\includegraphics[width=18cm]{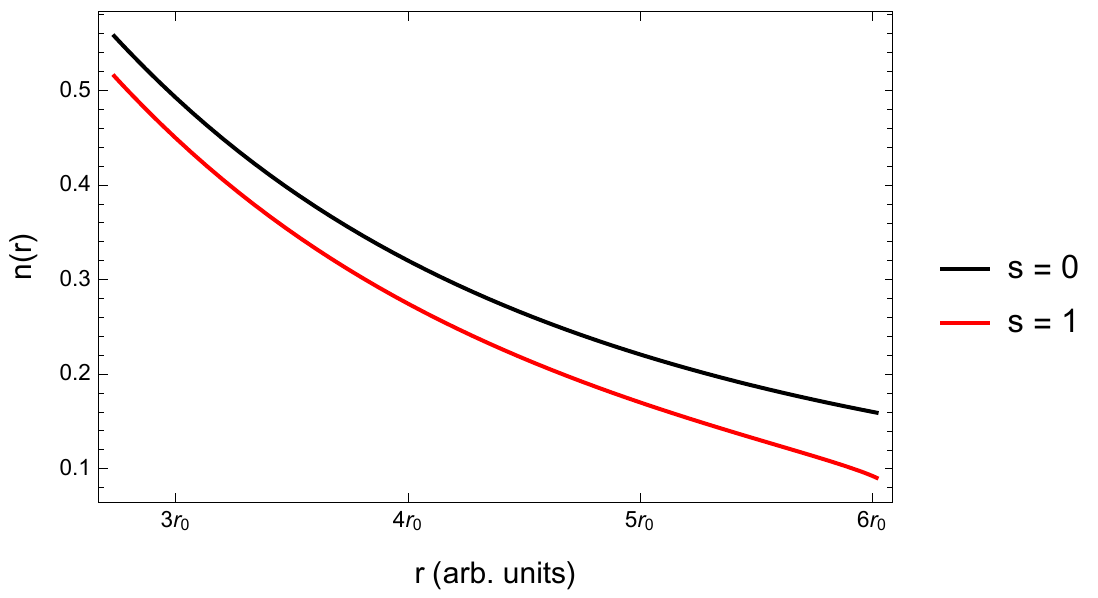}
\caption{Maxwell fish eye refraction index profile for $s=0$ and $s=1$ when $n_0 = 1.5$, $\lambda_0 = 1$, $r_0 = 5$.}
\end{center}
\end{figure}

%

\subsection{Trajectories}
\label{sec:fish-eye-polarized-trajectories}
Let us study the ray trajectories of the polarized light propagating in the medium with above constructed profile \eqref{spinFishEye0}.
One can see that
\be
\mathbf{r}\cdot\mathbf{J}=s\frac{\mathbf{rp}}{p},\quad \mathbf{r}\cdot\mathbf{T}_s = \frac{n_0}{\lambdabar_0}\frac{\mathbf{rp}}{p}\qquad\Rightarrow \qquad \mathbf{r}\cdot\left(\mathbf{J}-
\frac{s\lambdabar_0}{n_0}\mathbf{T}_s\right)=0.
\ee
Hence, ray trajectories are orthogonal  to the axis
\be
\mathbf{E}_3=\mathbf{J}-\frac{s\lambdabar_0}{n_0}\mathbf{T}_s,
\ee
and, therefore the trajectories belong  to the plane spanned by the following vectors
\be
\mathbf{E}_1=\mathbf{T}_s\times\mathbf{J},\qquad \mathbf{E}_2=\mathbf{E}_3\times\mathbf{E}_1=\Big(\mathbf{J^2}-s^2\Big)\left(\mathbf{T}_s -  \frac{4 s\lambdabar_0\kappa}{n_0}\mathbf{J}\right)
 \quad :\quad \mathbf{E}_3\cdot\mathbf{E}_2=\mathbf{E}_3\cdot\mathbf{E}_1=0.
\label{eee}
\ee
Then, from the expression $\mathbf{J}\cdot\big(\mathbf{r}\times\mathbf{T_s}\big)$ we  immediately obtain the solution  for the ray trajectories:
\be
\mathbf{r}\cdot\big(\mathbf{T_s}\times\mathbf{J}\big)=\Big(J^2-s^2\Big) \Big(2 - \frac{n_0}{n^{s}_{mfe}}\Big).
\label{traj}
\ee
This prompts us to  introduce the following orthogonal frame
\be
\mathbf{e}_i=\frac{\mathbf{E}_i}{|\mathbf{E}_i|}\; :\quad  \mathbf{e}_i\cdot\mathbf{e}_j=\delta_{ij},
\ee
where
\be
|\mathbf{E}_1|^2=\left(\mathbf{J}^2-s^2\right)\left(\frac{n^2_0}{\lambdabar^2_0}-{4\kappa \mathbf{J}^2}\right),\qquad |\mathbf{E}_3|^2=\left(\mathbf{J}^2-s^2\right)\left(1-\frac{4 s^2\lambdabar^2_0\kappa}{n^2_0}\right),\qquad |\mathbf{E}_2|^2=|\mathbf{E}_1|^2|\mathbf{E}_3|^2
\label{emod}\ee
Decomposing $\mathbf{r}$ over this frame, we introduce the  polar coordinates
\be
\mathbf{r}=x_1\mathbf{e}_1+x_2\mathbf{e}_2,\qquad x_1=r\cos\varphi,\quad x_2=r\sin\varphi.
\ee
Then,
having in mind Eqs. \eqref{eee} and \eqref{emod}, we can immediately rewrite  the  equation \eqref{traj} in polar coordinates
\be
1-  |\kappa||\mathbf{a}_s|r\cos\varphi=\frac{1+\kappa r^2}{1+\sqrt{1-\frac{4\kappa s^2\lambdabar^2_0}{n^2_0}(1+\kappa r^2)}},
\ee
where
\be
|\mathbf{a}_s|^2:=R^2_s-\frac 1\kappa,\qquad  R^2_s:=\frac{n^2_0-4\kappa s^2\lambdabar^2_0}{4\lambdabar^2_0\kappa^2(J^2-s^2)}.
\label{attr}\ee
So the trajectories of polarized light are not circles anymore, in contrast to the case of scalar waves. However, they can be attributed by the parameters ${\bf a_s}$ and $R_s$  \eqref{attr} which, in the limit $s\to 0$, become the center coordinate and the radius of the circle, respectively. Indeed, for $s=0$  the solution \eqref{traj} results in the   the equation for circle with the center located at $\mathbf{e}_1$ axis
\be
s=0\; :\quad \mathbf{r}\cdot\big(\mathbf{T}\times\mathbf{L}\big)=L^2(1-\kappa r^2) \quad\Rightarrow\quad \left(\mathbf{r}-\mathbf{a}_0\right)^2=R^2_0,
\ee
where
  \be \mathbf{a}_0:=\frac{\mathbf{T}\times\mathbf{L}}{2\kappa L^2},\quad |\mathbf{a}_0|^2=R^2_0-\frac{1}{\kappa},\quad R_0:=\frac{n_0}{2|\kappa|\lambdabar_0 L}.
\ee
\begin{figure}
\centering
\includegraphics[width=16cm]{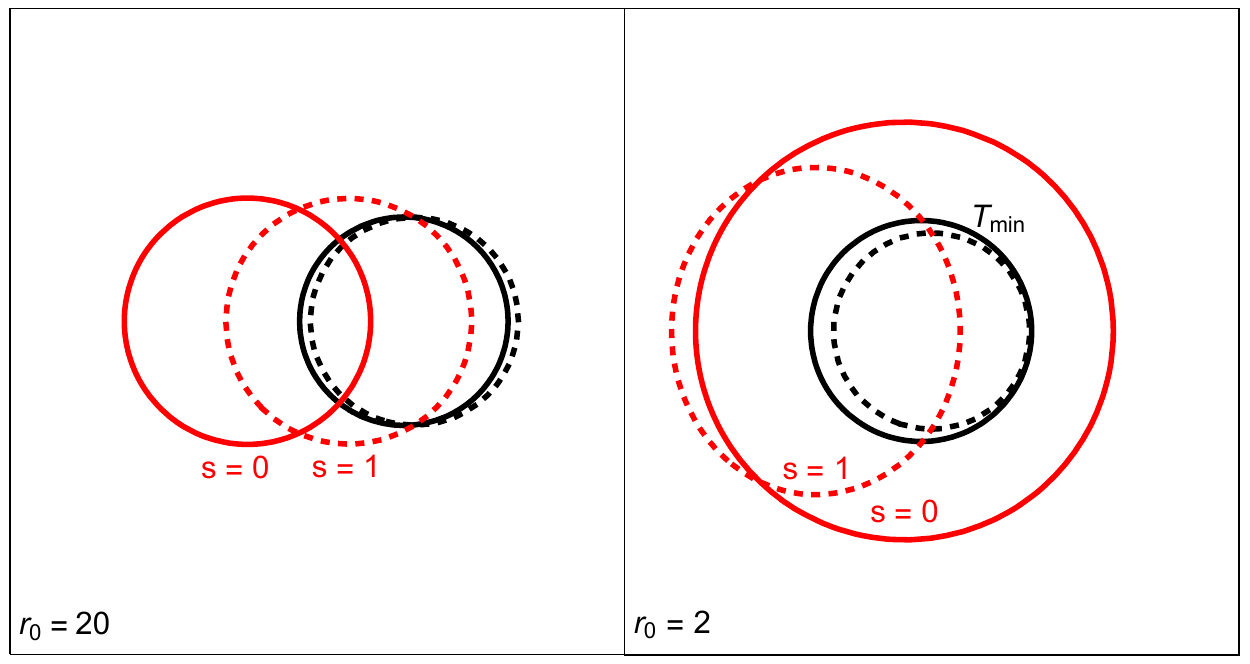}
\caption{Deformations of the ray trajectories for different values of $r_0$ when $n_0 = 1.5$, $\lambdabar_0 = 1$. The black curves correspond to the basic trajectories where $T = T_{min} = s/r_0$. The red (light gray) curves correspond to trajectories with intermediary value of $T$. Dashed curves are the trajectories corresponding to the same value of $T$ but for circularly polarized light ($s = 1$). The first figure ($r_0 = 20$) corresponds to the case when the deformations of the profile only result to the shift of the centers of the trajectories not affecting their shapes. Conversely, in the second figure ($r_0 = 2$), the deformations of the Maxwell fish eye result to highly deformed trajectories.}
\label{trajectory-deformations}
\end{figure}
Given $s\lambdabar_0\ \ll n_0/\kappa$, the deformation of circle is negligible, which is not the case for $s\lambdabar_0\  \sim n_0/\sqrt{\kappa}$.

As we can see from Fig. \ref{trajectory-deformations}, for $r_0 = 20$ the only notable manifestation of the polarization is the shift of the center of the trajectory.
However, it is worth noting that since the dashed lines are not circles anymore when talking about the center of the trajectory for  $s = 1$, we refer to the point which becomes the center of the circular trajectory when we pass from $s = 1$ to $s = 0$.
The second picture illustrates the circular trajectories and their deformations for  $s\lambdabar_0\  \sim n_0/\sqrt{\kappa}$. In this case,  the original  profile \eqref{Mfe0} and the deformed one  \eqref{spinFishEye0} differ drastically. The circular trajectories are notably deformed.

{  Detailed knowledge of trajectory parameters \eqref{attr} can be used in different applications. For example, in the conform mapping scheme the cloaking area is the outer space of closed trajectories \cite{Leonhardt06}. Therefore as it follows from \eqref{attr} there is no cloaking for polarized photons when $J\to s$.}

\subsection{Luneburg profile for polarized light}
\label{sec:fish-eye-polarized-luneburg}
Another interesting illustration of the scheme, proposed at the end of the first section, is the duality between the refractive profile corresponding to the well-known Luneburg lens (see Eq. \eqref{lun}) and the Hamiltionian of the harmonic oscillator.
\begin{equation}
n_{Lun}(\mathbf{r}) = n_0 \sqrt{2 - \left(\frac{\mathbf{r}}{r_0}\right)^2}.
\label{lun}
\end{equation}
Assume we have the Hamiltonian of the harmonic oscillator in presence of the magnetic monopole. So, we have $V(\mathbf{r}) = \omega^2\mathbf{r}^2/2$ and hence the Hamiltonian reads
\be
\label{oscHam}
H = \frac{p^2}{2} +\frac{\omega^2r^2}{2} + \frac{s^2}{2r^2}
\ee
As described in the first section, after fixing the energy surface $H=E$ of the Hamiltonian \eqref{oscHam} and performing the canonical transformation \eqref{ct} we arrive to the following equation:
\be
\label{oscHamFixedE}
r^2 + \omega^2p^2 + \frac{s^2}{p^2} - 2E = 0
\ee
By solving the equation \eqref{oscHamFixedE} in terms of $p$, we interpret the solution as the deformed refractive index profile in the optical Hamiltonian. Introducing the  notations
\be
r^2_0 := E  ,\qquad n_0 :=  \frac{r_0}{\omega},
\label{gammaE2}
\ee
we arrive to the final form of the deformed Luneburg profile
 \be
n^{s}_{Lun}(\mathbf{r})= \sqrt{\frac{1}{2} \left( n^{2}_{Lun} + \sqrt{n^{4}_{Lun} - \frac{4s^2 n^2_0}{r^2_0}}\right)}.
\label{spinLuneburg2}
\ee
Obviously, the above presented expression for the deformed Luneburg profile in the limit $s\rightarrow0$ transforms to Eq.\eqref{lun}.
\begin{figure}
\centering
\includegraphics[width=16.0cm]{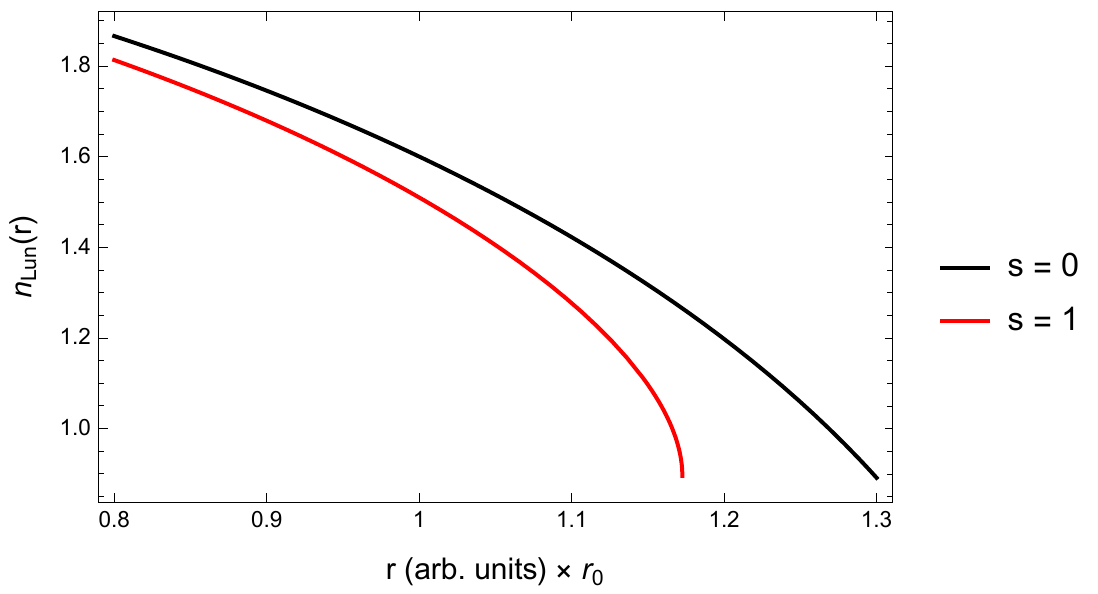}
\caption{Luneburg refraction index profile for $s=0$ and $s=1$ when $n_0 = 1.5$, $r_0 = 5$.}
\end{figure}

\subsection{Concluding remarks}
\label{sec:extended-symmetries-conclusion}
The standard Maxwell fisheye profile does not ensure closed ray trajectories for polarized photons (the only exception are linearly polarized photons corresponding to the $s=0$ spin stat), while the closeness of trajectories is the main property that is used in perfect imaging and cloaking phenomena.
In this paper, we suggested the deformation of the Maxwell fisheye profile which ensures the closeness of the trajectories of the ray trajectories for the polarized photons. We examined the properties of the deformed profile and have shown that the main difference between the cases of polarized and non-polarized photons is observed at the vicinity of wave and geometrical optics border $s\lambda_0/r_0\sim 1$.

Proposed modification scheme is applicable  for any isotropic refraction index $n(r)$. Namely, to preserve the qualitative properties of scalar wave trajectories for the propagating polarized light, we should replace it with the modified index $n^s(r)$ which is the solution
(with respect to $p$) of the following equation:
\be
p=\frac{1}{\lambdabar_0} n\left(\sqrt{r^2+\frac{s^2}{p^2}}\right),\quad\Rightarrow\quad p=n^s(r),
\ee
where $s$ is polarization of light.
The proposed deformation preserves the additional symmetries of the system (if any), and thus, guarantees the closeness of trajectories of polarized light.

Seemingly, the suggested scheme could be extended to some non-isotropic, but integrable profiles as well. On the other hand, non-isotropic integrable profiles are not common objects in the present study, though they obviously can be constructed by the use of existing integrable models.  For example,  choosing a textbook integrable system, the two-center Coulomb problem \cite{arnold} and performing trivial canonical transformation \eqref{ct} we can construct (taking into account the  expressions for constants of motion, see, e.g. \cite{2C}) anisotropic profile which could be interpreted as a superposition of two ``Maxwell fish eye" profiles. Furthermore, using the proposed scheme,  we can construct ``polarized Maxwell double fish-eye" profile as well, starting from the ``two-center MICZ-Kepler problem" \cite{kno}, i.e. from the two-center Coulomb problem specified by the presence of magnetic monopoles located at the attraction  centers.
We hope to consider this problem elsewhere.

\section[Discrete spectrum Radiation from a charged particle moving in a medium with Maxwell fish eye refraction index profile]
    {\textbf{Discrete spectrum Radiation from a charged particle moving in a medium with Maxwell fish eye refraction index profile}}
\label{sec:mfe-radiation}

\subsection{Introduction}
\label{sec:mfe-radiation-intro}

Recently the interest in the Maxwell fish eye \cite{bowolf} refraction index profile has increased dramatically. The reasons are its possible use in cloaking phenomena \cite{Leonhardt06}, perfect imaging \cite{Leonhardt09,Philbin10,Blaikie10,pazynin15}, quantum optics with single atoms and photons \cite{perczel}, optical resonators \cite{turks14,gevdav20}, etc. Earlier we have shown \cite{gevdav20} that apart from the spherical symmetry, Maxwell fish eye possesses an additional symmetry also. The extended symmetry leads to additional integrals of motion. In the geometrical optics limit, all the photon trajectories are closed and their parameters are expressed through the integrals of motion \cite{gevdav20}.
To detect a cloak in the Maxwell fish eye medium, it was suggested \cite{singap} to use the radiation induced by the motion of a charged particle.
Consideration was realized by exploring the dyadic Green's functions \cite{dyadic}. It was revealed that the emitted radiation is a mix of Cherenkov and transition radiations.

 As apposed to the research regarding light propagation in Maxwell fish eye medium, much less attention has been paid to light generation problems in that particular medium. However, it turns out that radiation emitted by a charged particle when it passes through such a medium possesses unique properties as well (see below).
In the present paper, we consider the spectrum and angular distribution of radiation from a charged particle moving in a Maxwell fish eye refraction profile medium. Instead of dyadic Green's function, we utilize the exact Green's function of scalar Helmholtz equation \cite{demkov71,poland,pazynin}. This approach allows us to obtain complete analytical expressions for radiation intensity that reveal new physical results.


\subsection{Initial Relations}
\label{sec:mfe-radiation-initial-relations}

We  start from the Maxwell equations for the field Fourier components
\begin{eqnarray}
 {\bf \nabla\times E(r,\omega)}=\frac{i\omega}{c}{\bf B(r,\omega)}, \quad {\bf \nabla\times H(r,\omega)}=\frac{4\pi}{c}{\bf j(r,\omega)}-\frac{i\omega}{c}{\bf D(r,\omega)}, \nonumber \\
 {\bf \nabla\cdot D}=4\pi \rho({\bf r},\omega),\quad {\bf \nabla\cdot B}=0,\quad {\bf B(r,\omega)}=\mu({\bf r},\omega) {\bf H(r,\omega)},\quad {\bf D(r,\omega)}=\varepsilon({\bf r},\omega){\bf E(r,\omega)}
\label{maxwell}
\end{eqnarray}
where $\rho$ and ${\bf j}$ are the charge and current densities associated with the moving particle.  We proceed with the calculations using vector potential ${\bf A}$ and scalar potential $\phi$ instead of ${\bf E}$ and ${\bf B}$. From Eq.(\ref{maxwell}) those can be introduced in the following way
\begin{equation}
{\bf B}={\bf \nabla\times A},\quad {\bf E}-\frac{i\omega}{c}{\bf A}={\bf \nabla}\phi
\label{potential}
\end{equation}
 By substituting the expressions for $E$ and $B$ into the second equation in Eq.(\ref{maxwell}), we get to the following equation
\begin{equation}
{\bf\nabla}\times\frac{1}{\mu}{\bf\nabla}\times{\bf A}=\frac{4\pi}{c}{\bf j}-\frac{i\omega}{c}\varepsilon({\bf\nabla}\phi+\frac{i\omega}{c}{\bf A})
\label{mixed}
\end{equation}
From Eq.(\ref{mixed}), assuming that $\mu({\bf r},\omega)$ is a slowly varying function in the space and using the property of double curl, we obtain
\begin{equation}
\nabla^2{\bf A}+\frac{\omega^2}{c^2}\varepsilon({\bf r})\mu({\bf r}){\bf A}=-\frac{4\pi}{c}\mu{\bf j}+{\bf\nabla}(\nabla\cdot{\bf A})+\frac{i\omega}{c}\varepsilon({\bf r})\mu({\bf r}) {\bf \nabla}\phi
\label{mixed2}
\end{equation}
 We take the gauge condition for inhomogeneous medium in the the following form
\begin{equation}
{\bf\nabla}(\nabla\cdot{\bf A})+\frac{i\omega}{c}\varepsilon({\bf r})\mu({\bf r}) {\bf \nabla}\phi=0
\label{gauge}
\end{equation}
and eventually find the equation for the vector potential
\begin{equation}
\nabla^2{\bf A}+\frac{\omega^2}{c^2}\varepsilon({\bf r})\mu({\bf r}){\bf A}=-\frac{4\pi}{c}\mu{\bf j}
\label{waveA}
\end{equation}
 Note that although the condition (\ref{gauge}) is similar to the Lorenz gauge condition for inhomogeneous media, it does not result to completely decoupled equations for vector and scalar potentials. However, our choice of the gauge condition leads to a less complex equation for the vector potential, which is more important when examining problems regardings radiation.
 Note also that when deriving Eq.(\ref{waveA}) we assume the slow variance only for $\mu({\bf r})$ but not for $\varepsilon({\bf r})$. This means that our consideration is correct for quite large class of materials (particularly all nonmagnetic mediums $\mu=1$). Also note that the form of equation for ${\bf A}$ can be changed depending on the gauge condition we choose \cite{bound}. The point is that in an inhomogeneous medium Lorenz gauge condition acquires different forms \cite{bound}. Again here we choose the one that leads to the simplest equation for the vector potential.

\begin{figure}
\begin{center}
\includegraphics[width=16cm]{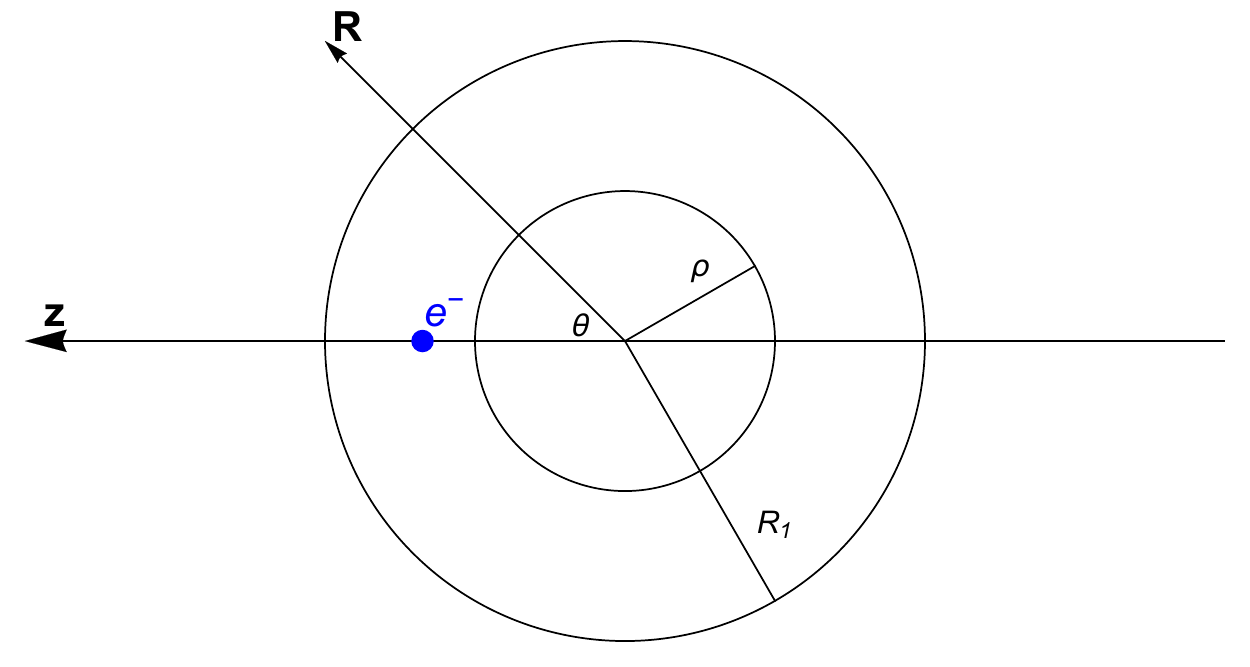}
\caption{Geometry of the problem. Observation point is far away from the charge and from the core of the refraction profile.}
\label{mfe-radiation-geom}
\end{center}
\end{figure}

It follows from the Eq.(\ref{waveA}) that  the radiation vector potential associated with the external source is directed similar to current density ${\bf j}$ which is assumed to be directed along $z$, see Fig.\ref{mfe-radiation-geom}. Therefore the radiation potential can be expressed through the Green's function of scalar Helmholtz equation
\begin{equation}
A_{zr}({\bf R})=-\frac{4\pi}{c}\int d{\bf r}G({\bf R,r})\mu({\bf r})j_z({\bf r}).
\label{radfield}
\end{equation}
The latter equation represents a particular solution of Eq.(\ref{waveA}) associated with an external source. To obtain the general solution, one should add also the solution of homogeneous equation. However, when examining the far field radiation, it is the particular solution Eq.(\ref{radfield}) that gives the main contribution.
Green's function satisfies the equation
\begin{equation}
[\nabla^2+\frac{\omega^2}{c^2}n^2({\bf R})]G({\bf R,r})=\delta({\bf R}-{\bf r})
\label{grfunc}
\end{equation}
where  $n({\bf r})=\sqrt{\varepsilon({\bf r})\mu({\bf r}})$ is the refraction index of the medium. We choose it in the form (see Fig.\ref{mfe-radiation-geom})
\begin{eqnarray}
n(r)=\frac{2n_0\rho^2}{r^2+\rho^2},\quad r<R_1 \nonumber\\
      1,\quad r>R_1
\label{fisheye}
\end{eqnarray}

Green's problem Eq.(\ref{grfunc}) is exactly solved \cite{demkov71,poland}
\begin{eqnarray}
G_{\nu}({\bf R,r})=-\frac{1}{4\pi\cos(\pi\nu)}\frac{\sqrt{(R^2+\rho^2)(r^2+\rho^2)}}{|{\bf R-r}|\sqrt{R^2r^2+2\rho^2{\bf Rr}+\rho^4}}\times \nonumber\\\times \sin\left[(\nu+1/2)\arccos\left(-1+\frac{2\rho^2({\bf R-r})^2}{(R^2+\rho^2)(r^2+\rho^2)}\right)\right]
\label{grfunsol}
\end{eqnarray}
where
\begin{equation}
\nu=\frac{-1+\sqrt{1+4n_0^2k^2\rho^2}}{2}.
\label{nu}
\end{equation}
Here $k=\omega/c$ and $\nu\neq m+1/2, -m-3/2$ $(m\in \textrm{N})$. At these specific values, as it is seen from Eq.(\ref{grfunsol}), Green's function is divergent and one needs another expression \cite{poland}
\begin{eqnarray}
\tilde{G}_{m+1/2}({\bf R,r})=(-)^m\frac{1}{4\pi^2}\frac{\sqrt{(R^2+\rho^2)(r^2+\rho^2)}}{|{\bf R-r}|\sqrt{R^2r^2+2\rho^2{\bf Rr}+\rho^4}}\times\nonumber\\  \times\left\{\cos\left[(m+1)\arccos\left(-1+\frac{2\rho^2({\bf R-r})^2}{(R^2+\rho^2)(r^2+\rho^2)}\right)\right]\times \right.\nonumber\\ \left.\times\arccos\left(-1+\frac{2\rho^2({\bf R-r})^2}{(R^2+\rho^2)(r^2+\rho^2)}\right)+\right.\nonumber\\
\left. +\frac{\sin\left[(m+1)\arccos\left(-1+\frac{2\rho^2({\bf R-r})^2}{(R^2+\rho^2)(r^2+\rho^2)}\right)\right]}{2(m+1)}\right\}
\label{grfuncother}
\end{eqnarray}

The current density corresponding to the particle with charge $e$  moving along the $z$ axis with velocity $v$ has the form  ${\bf j}({\bf r}, t)=e{\bf v}\delta(x)\delta(y)\delta(z-v_z t)$. The corresponding Fourier component which determines the Fourier component of the radiation vector potential will have the following form
\begin{equation}
{\bf j}({\bf r},\omega)=\frac{e{\bf v}}{v}\delta(x)\delta(y)e^{ik_0z}
\label{chargecurr}
\end{equation}
where $k_0=\omega/v$. Using the expressions for Green's functions Eqs.(\ref{grfunsol},\ref{grfuncother}) and the expression for the current density Eq.(\ref{chargecurr}), one can find radiation potential and radiation intensity.

\subsection{Radiation potential}
\label{sec:mfe-radiation-potential}

First, we determine the radiation potential in the region $R<R_1$
\begin{equation}
A({\bf R})=-\frac{4\pi}{c}\int_{r<R_1}d{\bf r}G({\bf R,r})\mu({\bf r})j({\bf r}),\quad R<R_1
\label{vecpot}
\end{equation}
Here $A\equiv A_z, j\equiv j_z$ .
According to the Green's theorem, in Eq.(\ref{vecpot}) it should also appear an additional surface integral over the sphere of radius $R_1$. However, for large $R_1\gg \rho$, this surface term falls  faster than $1/R_1$, therefore it does not give a contribution to the radiation potential. As it seen from Eq.(\ref{grfunsol}) Green's function has singularities at the points $\nu=m+1/2$. In order to overcome this difficulty we assume a small imaginary part for $n_0$ and correspondingly for $\nu$. As it is noted in \cite{poland} the expression Eq.(\ref{grfunsol}) is correct for complex values of $\nu$  either.  For these discrete values $\nu$ the integral Eq.(\ref{vecpot}) can be calculated analytically \cite{ryzhik}. We present the results for $\nu=1/2+iIm[\nu_{1/2}]$ and $\nu=3/2+iIm[\nu_{3/2}]$, $Im[\nu]\ll 1$
\begin{eqnarray}
A_{1/2}(R)=-\frac{4e}{c}\frac{i sign(v)K_0(k_0\rho)}{\sinh(\pi Im[\nu_{1/2}])}\frac{\rho}{\sqrt{R^2+\rho^2}},\nonumber\\
A_{3/2}(R)=-\frac{8ie sign(v)}{c \sinh(\pi Im[\nu_{3/2}])} \frac{\rho}{\sqrt{R^2+\rho^2}}\left[(k_0\rho)K_0(k_0\rho)-K_1(k_0\rho)\right],\quad R<R_1
\label{vecpotspec}
\end{eqnarray}
where $K_{0,1}$ are the modified Bessel functions of second kind \cite{ryzhik} and non-magnetic medium ($\mu\equiv 1$) is assumed (see below).
When obtaining Eq.(\ref{vecpotspec}) in the limit $R,R_1\gg \rho$ we extend the integral limits in Eq.(\ref{vecpot}) to infinity and neglect all the terms smaller $1/R$. Note that radiation potential depends only on the module of the vector ${\bf R}$. Non isotropic terms are possible for non-discrete values of $\nu$, however they are small in terms of the parameter $1/R$. In order to find the radiation intensity, one should know the radiation fields in the vacuum region that matches the solutions given in Eq.(\ref{vecpotspec}).

\begin{figure}
\begin{center}
\includegraphics[width=17cm]{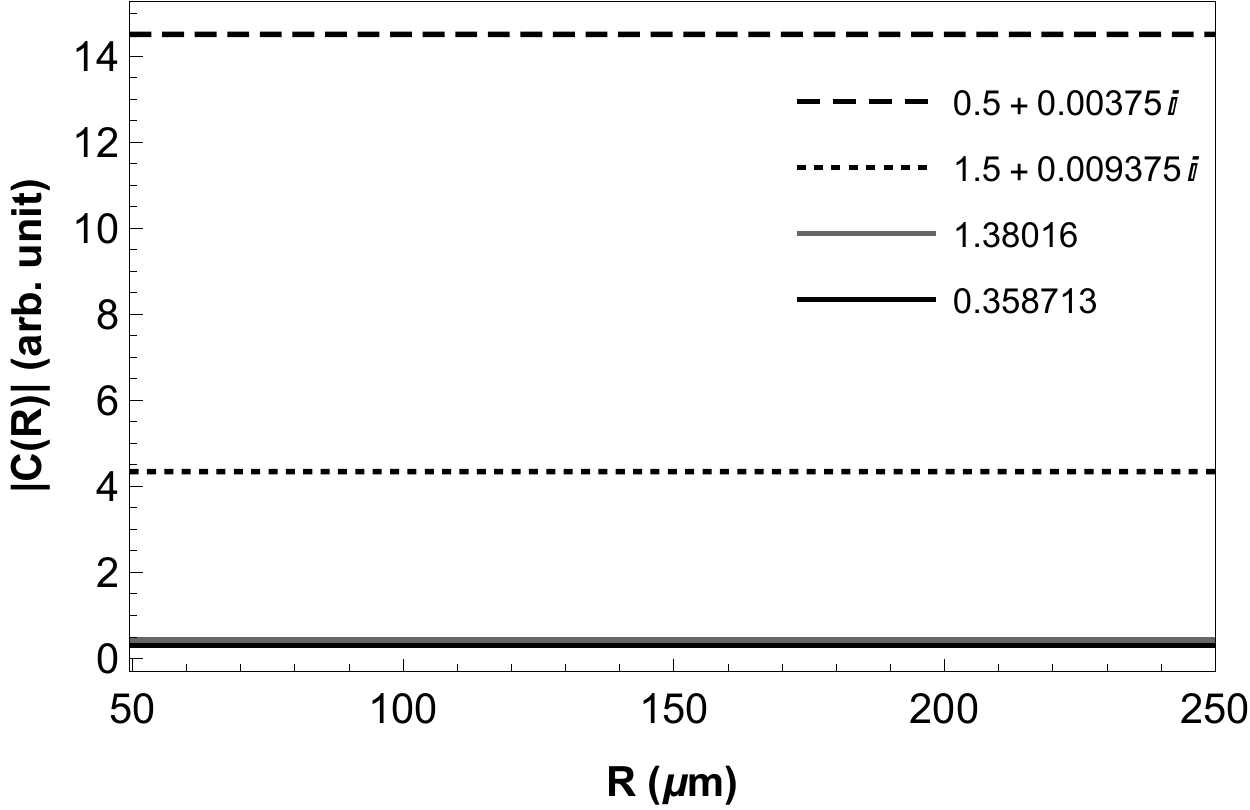}
\caption{$|C(R)|$ dependence on $R$ for different values of $\nu$ when $n_0 = 5, \beta = 0.9$. Straight lines mean that $|A(R)|\sim 1/R$. The amplitude of the radiation is negligible for $\nu \neq m+1/2$.}
\label{fig-spectrum-new}
\end{center}
\end{figure}

In the absence of external sources, the isotropic solution of Eq.(\ref{waveA}) is chosen in the form
\begin{equation}
A_v(R)=C\frac{e^{ikR}}{R},\quad R>R_1
\label{vac}
\end{equation}
The constant $C$ should be found from the boundary conditions. Since there is no any current on the sphere $R_1$ (${\bf R_1}\neq \hat{\bf z}R_1$), the magnetic field is finite and correspondingly we have (see for example \cite{bound})
\begin{equation}
{\bf nxA_1=nxA_2}
\label{bound}
\end{equation}
where ${\bf n}$ is unit vector normal to the boundary surface. In our case, this leads to the equation $A_v(R_1)=A_{1/2}(R_1)$. Using Eqs.(\ref{vecpotspec},\ref{vac}) one finds
\begin{equation}
C_{1/2}(R_1)=-\frac{4iesign(v)}{c\sinh(\pi Im[\nu_{1/2}])}\rho K_0(k_0\rho)e^{-ikR_1}
\label{constant}
\end{equation}
When obtaining Eq.(\ref{constant}) we assumed that $R_1\gg\rho$. One can also find $C_{3/2}$ and other values of $C$ for $Re[\nu]=m+1/2$ in analogous manner. For the non-discrete values of $\nu$, $C$ can be found numerically (see Fig.\ref{fig-spectrum-new}).

As it follows from Fig.\ref{fig-spectrum-new} one can expect any significant radiation emission only at frequencies for which $\nu=m+1/2$.

\subsection{Radiation intensity}
\label{sec:mfe-radiation-intensity}

To find the radiation intensity, one should know the electric and magnetic fields far away from the charge outside of the medium area $R>R_1\gg\rho$ (see Fig.\ref{mfe-radiation-geom}).
Recall that the radiation vector potential is directed along the $z$ axis and ${\bf A}\equiv A_z$. Therefore from Eq.(\ref{vac}) the magnetic field ${\bf B}={\bf \nabla\times A}$ at the observation point $R$ is given by expressions

\begin{equation}
B_x({\bf R})=C\frac{ikR_ye^{ikR}}{R^2},\quad B_y({\bf R})=-C\frac{ikR_xe^{ikR}}{R^2},\quad B_z\equiv 0
\label{magfield}
\end{equation}

Here we keep only the terms which are proportional to $O(1/R)$ that give contribution to the radiation intensity. The magnetic field energy density follows from Eq.(\ref{magfield})
\begin{equation}
U_B=\frac{|{\bf B}|^2}{8\pi}=|C|^2\frac{k^2\sin^2\theta}{8\pi R^2}
\label{mageng}
\end{equation}
The electric field energy density in the vacuum is equal to the magnetic field energy density and the radiation intensity is determined through the electromagnetic field energy density as $I(\theta)=cR^2U$, where $U=2U_B$. Using Eqs.(\ref{constant},\ref{magfield}) for the radiation intensity
at $Re[\nu]=1/2$, we have
\begin{equation}
I_{1/2}(\theta)=\frac{4e^2}{\pi c}\frac{k^2\rho^2K_0^2(k_0\rho)}{\sinh^2(\pi Im[\nu_{1/2}])}sin^2\theta
\label{inten}
\end{equation}
It follows from Eq.(\ref{nu}) that for $n_0^2=\varepsilon+iIm[\varepsilon],\quad Im[\varepsilon]\ll \varepsilon$ and $\nu=1/2+iIm[\nu_{1/2}]$
\begin{equation}
k\rho=\frac{\sqrt{3}}{2\sqrt{\varepsilon}},\quad Im[\nu_{1/2}]=\frac{3Im[\varepsilon]}{8\varepsilon}
\label{spcase}
\end{equation}

Substituting Eq.(\ref{spcase}) into Eq.(\ref{inten}),we finally obtain
\begin{equation}
I_{1/2}(\theta)=\frac{3e^2}{\pi c}\frac{K_0^2(\frac{\sqrt{3}}{2\beta\sqrt{\varepsilon}})}{\sinh^2(\frac{3\pi Im[\varepsilon]}{8\varepsilon})}sin^2\theta
\label{final}
\end{equation}
where $k=k_0\beta$ and $\beta=v/c$. Here we present a peak intensity corresponding to the wavelength $\lambda=4\pi\sqrt{\varepsilon/3}\rho$ $(Re[\nu]=1/2)$.  Similar expressions can be written for smaller peak intensities $Re[\nu]=3/2$, etc. Intensities for non-discrete frequencies are significantly smaller (see Fig.\ref{fig-spectrum-new}). It is well known \cite{ryzhik} that modified Bessel function $K_0$ is exponentially small for large values of the argument. Therefore from Eq.(\ref{final}) we can state that for the existence of radiation, the following condition should be satisfied
\begin{equation}
\beta>\frac{\sqrt{3}}{2\sqrt{\varepsilon}}
\label{cheren}
\end{equation}
This is the analogue of Cherenkov condition \cite{mik72} for the Maxwell fish eye profile. Note that the radiation considered here is the mix of Cherenkov and transition radiations, see also \cite{singap}. As it follows from Eq.(\ref{cheren}), radiation emission condition in Maxwell fish eye profile is weaker than the ordinary Cherenkov condition in the homogeneous medium  with refraction index $\sqrt{\varepsilon}$, $\beta>1/\sqrt{\varepsilon}$. However, it is stronger than the Cherenkov condition for the homogeneous medium with refraction index $2\sqrt{\varepsilon}$. It is interesting that condition obtained for totally inhomogeneous medium is very similar to Cherenkov condition for the homogeneous medium.

 As it seen from Eq.(\ref{final}), the angular distribution of the intensity is like that of dipole radiation. Moreover, the maximum intensity is reached in the directions normal to the particle velocity.
Note that the radiation intensity does not depend on the cutting distance $R_1$. This is the result of the fact that radiation fields behave as $1/R$ and correspondingly $|C|$ does not depend on $R_1$, see Eq.(\ref{constant}). However this is not always the case and in the lossless medium ($Im[\varepsilon]=0$) the situation is changed (see below).

At the end of this paragraph we present radiation intensity for the impedance match medium $\mu(r)=\varepsilon(r)=n(r)$ (see \cite{Philbin10})
\begin{equation}
I_{1/2}^i(\theta)=\frac{9e^2}{\pi c}\frac{K_1^2(\frac{\sqrt{3}}{2\beta\sqrt{\varepsilon}})}{\varepsilon\beta^2\sinh^2(\frac{3\pi Im\varepsilon}{8\varepsilon})}sin^2\theta
\label{imp}
\end{equation}

Comparing with the nonmagnetic medium case Eq.(\ref{final}) one can see that the main difference is the additional particle energy dependence ($\sim 1/\beta^2$).

\subsection{Lossless medium. Non isotropic radiation}
\label{sec:mfe-radiation-intensity-lossless}

\begin{figure}
\begin{center}
\includegraphics[width=16cm]{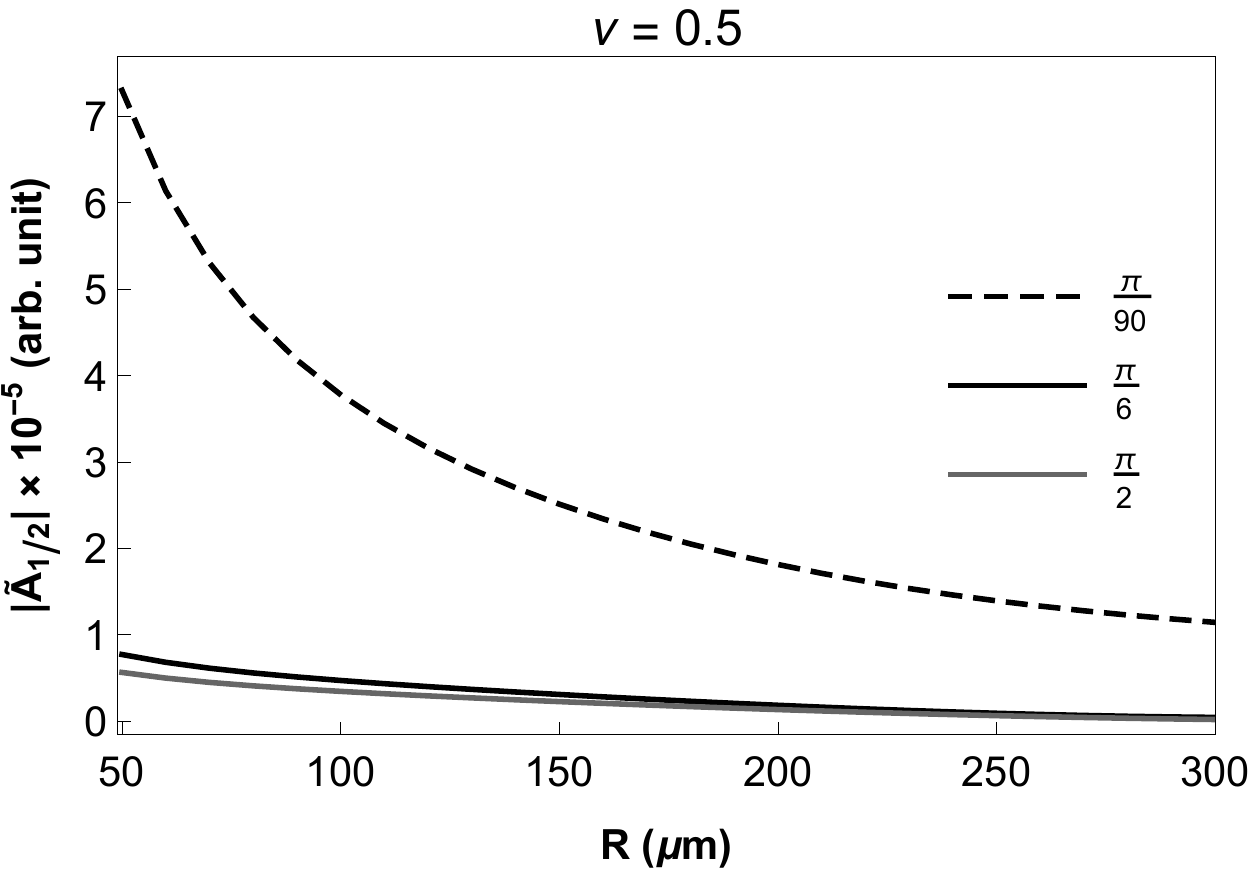}
\caption{R-dependence of radiation potential given by Eq.(\ref{anis}) for different angles. Here we take the observation point on the cutting boundary $R\equiv R_1$.}
\label{angular-distribution-versus-r}
\end{center}
\end{figure}

\begin{figure}
\begin{center}
\includegraphics[width=16cm]{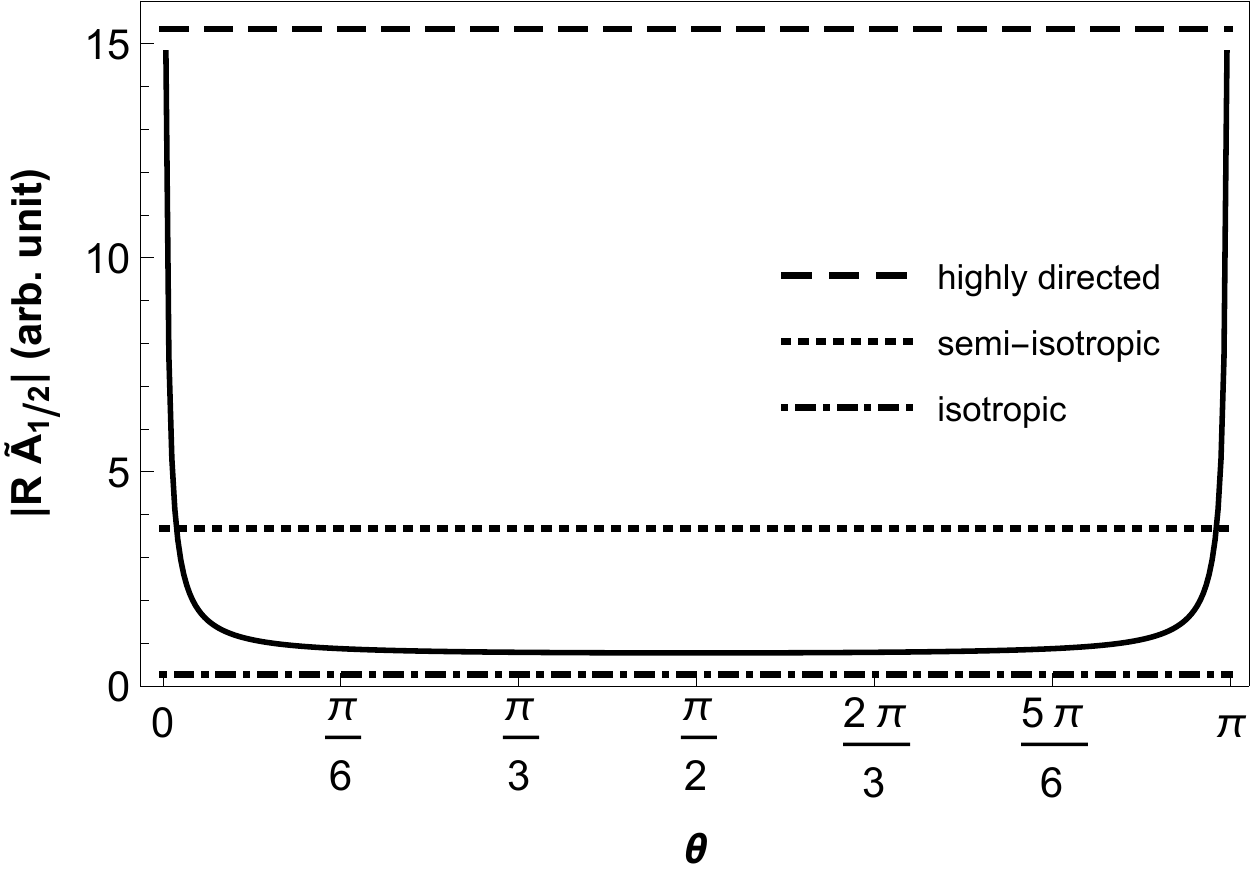}
\caption{Angular distribution of radiation vector potential.  Straight lines are the potentials for the media with losses for different values of refraction index imaginary part }
\label{angular-distribution}
\end{center}
\end{figure}

The main difference is happening at the discrete frequencies $\nu=m+1/2$. In this case, the radiation potential is determined through the generalized Green's function Eq.(\ref{grfuncother}). Our estimates show that the radiation potential at large distances $R\gg \rho$ behaves
as
\begin{eqnarray}
\tilde{A}_{1/2}({\bf R})&\sim \frac{8esign(v)}{\pi c}\int_{-R_1}^{R_1} dz\frac{(z^2-\rho^2-4\rho^2 z\cos\theta/R)e^{ik_0z}}{\sqrt{[(z-R\cos\theta)^2+R^2\sin^2\theta](z^2+\rho^2)[(z+\rho^2\cos\theta/R)^2+\rho^4\sin^2\theta/R^2]}} \nonumber\\
&\arccos\frac{\rho^2-z^2-4\rho^2 z\cos\theta/R}{z^2+\rho^2}
\label{anis}
\end{eqnarray}
Unfortunately this integral can not be taken analytically and we are forced to use some numerical estimations. Nevertheless we will try to collate analytic and numerical results and make qualitative predictions on radiation properties in different cases.
The second term in Eq.(\ref{grfuncother}) for $Re[\nu]=1/2$ has an isotropic character and is analogous to the regular case in Eq.(\ref{grfunsol}). It is obvious that at small angles $\theta\to 0$, the main contribution to the integral Eq.(\ref{anis}) gives the pole at $z=-\rho^2\cos\theta/R$. Conversely, the contribution of the pole $z=R\cos\theta$ at large distances is negligible because of the oscillations $e^{ik_0z}$ in the integral.
We present the results of numerical estimates of the integral Eq.(\ref{anis}) in Fig.\ref{angular-distribution-versus-r} and Fig.\ref{angular-distribution}.
As it follows from Fig. \ref{angular-distribution-versus-r} at large distances the potential behaves as $\sim 1/R$. However, the amplitude is different for different observation angles in contrary to the case in the previous paragraph.

The radiation potential in the lossless medium is highly anisotropic. It follows from Fig. \ref{angular-distribution-versus-r} that at large observation angles the potential is significantly smaller. Maximum potential is reached at small angles from the particle trajectory. Modifying the imaginary part of $n_0$, one can observe a transition from highly directed to isotropic radiation potential.  As it is shown in the previous paragraph, isotropic radiation potential leads to dipole like radiation intensity. However, in the lossless medium the radiation potential is highly directed.
Note that the $R$-dependence for all angles is $\sim 1/R$, see Fig.\ref{angular-distribution-versus-r} . Straight lines in Fig.\ref{angular-distribution} represent the radiation potentials  for different values of $Im[n_0]$ calculated via Eq.(\ref{vecpotspec}). The $U$-shape curve is obtained for the lossless medium using generalized Green's function Eq.(\ref{grfuncother},\ref{anis}). The actual radiation potential is determined by the curve below the straight lines.  Therefore one can distinguish three different regimes of radiation depending on the losses of the medium ($Im[n_0]$), see Fig.\ref{angular-distribution}.

Physically, the above-mentioned transition from one angular distribution to another can be understood in the following way. Vector potential of a moving charge in the vacuum behaves like $A_0(R)\sim K_0(\sqrt{k_0^2-k^2}R|\sin\theta|)$. This means that it is concentrated around the direction of the particle's velocity because for large arguments $K_0$ falls exponentially. This alone is not a photonic field yet. When one adds a nonhomogeneous medium these pseudophotons scatter becoming real photons. In the Maxwell fish eye medium without losses, photons are totally reflected from layers with decreasing refraction index. Because of the total internal reflection, those remain around the charge. When the losses are taken into account photons evanescently penetrate through the layers and eventually result in the isotropic angular distribution for large enough losses as seen in Fig.\ref{angular-distribution-versus-r}. This transition is similar to the phenomenon of attenuated total reflection (ATR), see, for example, \cite{atr}.

\subsection{Concluding remarks}
\label{sec:mfe-radiation-coclusion}
We have considered the radiation from a charged particle moving through a medium with Maxwell fish eye refraction index profile. The spectrum of radiation has discrete character. The main emitted wavelength is proportional to the radius of refraction profile $\lambda=4\pi\sqrt{\varepsilon/3}\rho$. In the regular medium (with losses) the radiation has dipole character, whereas in the lossless medium it is highly directed. In the intermediate regime with moderate losses radiation will be non-isotropic.

  So far we assumed that particle trajectory along $z$ passes through the origin. If the trajectory is at some distance $d$ from the origin then the corresponding current density is determined as
\begin{equation}
{\bf j}({\bf r},\omega)=e\frac{{\bf v}}{v}\delta(x-d_x)\delta(y-d_y)e^{ik_0z}
\label{impact}
\end{equation}
Similar calculations show that all expressions keep their form except $\rho$ in the argument of the Bessel function Eq.(\ref{inten}). It should be substituted by $\sqrt{\rho^2+d^2}$ where $d^2=d_x^2+d_y^2$. So for the main emitted wavelength $\lambda=4\pi\sqrt{\varepsilon/3}\rho$, one has
\begin{equation}
I_{1/2}^d(\theta)=\frac{3e^2}{\pi c}\frac{K_0^2(\frac{\sqrt{\frac{3}{4\varepsilon}+k^2d^2}}{\beta})}{\sinh^2(\frac{3\pi Im\varepsilon}{8\varepsilon})}sin^2\theta
\label{intendis}
\end{equation}
Correspondingly, Cherenkov condition acquires the form
\begin{equation}
\beta>\frac{\sqrt{3}}{2\sqrt{\varepsilon}}\sqrt{1+\frac{d^2}{\rho^2}}
\label{cherend}
\end{equation}
This condition means that the impact distance should be smaller than the emitted wavelength $d<\lambda/2\pi$.

 In the anisotropic case radiation potential depends not only the ratio $d/\lambda$ as in isotropic case but also $d/\rho$. Therefore here restriction on the impact parameter is stronger $d/\rho\ll 1$.

Note that Maxwell fisheye millimeter scale systems already exist in 2D \cite{2D} as well as in  3D \cite{3D}. Therefore they can be used for the generation of radiation in microwave and terahertz regions. As it seen from Eq.(19), radiation intensity does not depend on the cutting parameter $R_1$ (except the lossless medium case). This means that real systems with Maxwell fish eye profile can have a size of order radius $\rho$ \cite{2D,3D}.


\cleardoublepage
\phantomsection
\section*{Conclusion}

\addcontentsline{toc}{section}{Conclusion}

\label{sec:conclusion}
Let us summarize our results.

The first chapter is an introduction and some general concepts that are investigated through the dissertation. In this section we discuss various types of inhomogeneous media that are considered in this dissertation. Firstly, we give a brief overview on photonic crystals which are media with periodic inhomogeneity and result in many interesting phenomena such as polarization rotation, light straightening effect etc. Secondly we present the other type of media with symmetric inhomogeneity - the so called GRIN media. As opposed to photonic crystals, in GRIN media the refrative index is a continuous function of coordinate. Moreover, we briefly present the importance of specific refractive index profiles (Maxwell fish eye, Luneburg lense) in various phenomena such as the cloaking effect, perfect imaging etc.   

In \chap{\ref{sec:pol_rot}} we investiagate the resonance polarization change in photonic crystal. Firstly, we present the expermental set-up that was used to detect the polarization change. In the following two sections it is presented the theory which is based on Maxwell's equations with two dimensional inhomogeneous permittivity. In \cite{gevdav19} we have proposed the theory for the TE waves and obtained a reasonable correspondence between the theoretical and experimental resonant frequencies.

In \chap{\ref{sec:extended-symmetries}} we examine the additional symmetries of Maxwell Fish eye and have shown that ray trajectories in Maxwell fish eye are closed and the expressions for ray trajectories are found directly from the integrals of motion. Moreover, we have shown that there exists a photon state with maximal angular momentum and it can be used as a possible means for creating optical resonators. 

In \chap{\ref{sec:fish-eye-polarized}} we examine the propagation of polarized light in the medium with Maxwell fish eye refraction index profile. In \autoref{sec:fish-eye-polarized-scalar-waves}, we describe the Hamiltonian formulation of the geometric optical system given by the action \eqref{gactions2}. We also present some other textbook facts on the duality between Coulomb and free-particle systems on a (pseudo)sphere which allow us to relate the Maxwell fish eye and Coulomb profiles and will be used in our further consideration. In \autoref{sec:fish-eye-polarized-scalar-waves}, we present the Hamiltonian formalism for the polarized light propagating in an optical medium and propose the general scheme of the deformation of an isotropic refraction index profile which allows us to restore the initial symmetries after the inclusion of polarization. In \autoref{sec:fish-eye-polarized-spin-inclusion}, we use the proposed scheme for the construction of a “polarized Maxwell fish eye” profile which inherits all the symmetries of the original profile when light polarization is taken into account. We present the explicit expressions for the symmetry generators of the corresponding Hamiltonian system and find the expressions of the Casimirs of their symmetry algebra. In \autoref{sec:fish-eye-polarized-trajectories} the explicit expressions for the trajectories of polarized light are presented. It is shown that these trajectories are no longer orthogonal to the angular momentum but turn to a fixed angle relative to it. Despite deviations from circles, these trajectories remain closed. We show that light polarization violates the additional symmetries of the medium so that ray trajectories no longer remain closed. Afterwards we suggest a modified, polarization-dependent  Maxwell fish eye refraction profile which restores all the symmetries of the initial profile and yields closed trajectories of polarized light. Explicit expressions for the polarization-dependent integrals of motion and the solutions of corresponding ray trajectories are also presented.

In \chap{\ref{sec:mfe-radiation}} we consider radiation properties of the Maxwell fish eye. That is, we examine the radiation from a charged particle when is passes through the core of the Maxwell fish eye refraction index profile. Maxwell’s equations are used to describe the electromagnetic fields in an inhomogeneous medium taking into account the moving charge as an external source. Also, we have constructed the well-known Poynting’s vector which will be used to derive the radiation intensity. Whenever appropriate, numerical estimations of integrals were completed. We have shown that the radiation spectrum has a discrete character. The main emitted wavelength is proportional to the refractive profile's radius and has a dipole character in a regular medium. Cherenkov like threshold velocity was established. 
\newpage

\cleardoublepage
\phantomsection
\section*{Summary}
\addcontentsline{toc}{section}{Summary}
Here we present the outline of the main results of this thesis.

\begin{itemize}
	\item A new theory based on Maxwell equations has been proposed which explains the resonance polarization change in dilute photonib crystal
	\item Parameters of ray trajectories in Maxwell fish eye have been expressed through the integrals of motion
	\item We observed the existence of a photon state with maximal angular momentum, which can be used as an optical resonator
	\item Spin Hall effect in an extended symmetry profile has been predicted
	\item We have shown that polarization violates the additional symmetries of the medium so that ray trajectories no longer remain closed
	\item We suggest a modified, polarization-dependent Maxwell fish eye refraction index which restores
all symmetries of the initial profile and yields closed trajectories of polarized light
	\item Explicit expressions for the polarization-dependent integrals of motion and the solutions of corresponding ray trajectories are presented
	\item A generalized scheme has been proposed which allows to find the deformation of an arbitrary refraction index profile when spin is taken into account
	\item The deformation of the well-known Luneburg profile is also presented
	\item It is shown that the radiation from a charged particle moving in a medium with a Maxwell’s fish-eye refraction index profile has a discrete spectrum 
	\item The main emitted wavelength is proportional to the refractive profile’s radius and has a dipole character in a regular medium. A Cherenkov-like threshold velocity is established
	\item We have shown that there is a  cardinal rearrangement of angular distribution in a lossless medium caused by the total internal reflection in a lossless medium as opposed to photons’ attenuated
total reflection in the regular medium
	\item Lossless medium ensures that both directed and monochromatic emission can serve as a light source in the corresponding regions
\end{itemize}

\section*{Acknowledgements}
\label{sec:acknowledgements}

The success and final outcome of this dissertation required a lot of guidance and assistance from
many people. All that I have done is only due to such supervision and assistance and I will not
forget to express my appreciation to them.

First of all, I would like to express my gratitude to my academic advisor Prof. Zhirayr Gevorkian for his continuous support, mentorship and for the uncountably many things that I have learned during the Ph.D program. He motivated me to grow as a researcher and his ongoing support was of key importance in the sense of writing the thesis. He also included me in many grants and projects. The financial support from these grants allowed me to concentrate on my research. 

I would like to express my deepest appreciation to my advisor Prof. Armen Nersessian for his continuous guidance and support. Apart from the huge amount of knowledge that he shared, he also made it possible to cooperate with international community of theoretical physics and report the results in several international conferences.

I would like to thank Prof. Arsen Hakhumian for his ongoing support during the Ph.D program.

I would like to thank all the staff of Institute of Radiophysics and Electronics for their support, patience and encouragement. They have assisted me with any problem and question and backed me from the very first day I entered the institute. My special thanks to Tigran Zakaryan, Emil Asmaryan and Ashkhen Yesayan.

\cleardoublepage
\phantomsection

\end{document}